\documentclass[12pt]{article}
\pdfoutput=1
\usepackage{latexsym}
\usepackage{cite,epsfig,amssymb,euscript,slashed}
\usepackage[normalem]{ulem} 
\usepackage{amsmath}
\usepackage{float}
\usepackage{array,calc,epsfig}
\usepackage[linktocpage]{hyperref}
\usepackage{bbm}
\usepackage{braket}
\usepackage{xcolor}
\usepackage{fancybox}

\def\be{\begin{equation}}
\def\ee{\end{equation}}
\def\bseq{\begin{subequations}}
\def\eseq{\end{subequations}}

\def\bea{\begin{eqnarray}}
\def\eea{\end{eqnarray}}

\def\bseq{\begin{subequations}}
\def\eseq{\end{subequations}}

\numberwithin{equation}{section} 
\usepackage[]{graphicx}

\def\d {{\rm d}}

\def\ii           {{\rm i}}

\def\Re           {{\rm Re\hskip0.1em}}
\def\Im           {{\rm Im\hskip0.1em}}

\def\sqr#1#2{{\vcenter{\vbox{\hrule height.#2pt
 \hbox{\vrule width.#2pt height#1pt \kern#1pt \vrule width.#2pt}\hrule
 height.#2pt}}}}

\def\a{\alpha}
\def\b{\beta}

\def\d{\text{d}}

\def\slashchar#1{\setbox0=\hbox{$#1$}           
\dimen0=\wd0                                 
\setbox1=\hbox{/} \dimen1=\wd1               
\ifdim\dimen0>\dimen1                        
\rlap{\hbox to \dimen0{\hfil/\hfil}}      
#1                                        
\else                                        
\rlap{\hbox to \dimen1{\hfil$#1$\hfil}}   
/                                         
\fi}

\begin{document}
\font\cmss=cmss10 \font\cmsss=cmss10 at 7pt

\title{\bf ModMax meets Susy}

\author{Igor Bandos${}^{a,b}$\footnote{e-mail: {\tt  igor.bandos@ehu.eus }}, Kurt Lechner${}^{c,d}$\footnote{e-mail: {\tt  kurt.lechner@pd.infn.it }}, Dmitri Sorokin${}^{d,c}$\footnote{e-mail: {\tt  dmitri.sorokin@pd.infn.it }} and Paul K. Townsend$^e$\footnote{e-mail: {\tt pkt10@cam.ac.uk}}}

\date{}

\maketitle

\vspace{-1.5cm}

\begin{center}

\vspace{0.5cm}\textit{\small
${}^a$ Department of
	Theoretical Physics, University of the Basque Country UPV/EHU,\\
	P.O. Box 644, 48080 Bilbao
	\\ ${}^b$ IKERBASQUE, Basque Foundation for Science, 48011 Bilbao, Spain}

\vspace{0.3cm}
\textit{\small ${}^c$ Dipartimento di Fisica e Astronomia ``Galileo Galilei",  Universit\`a degli Studi di Padova \\
${}^d$ I.N.F.N. Sezione di Padova, Via F. Marzolo 8, 35131 Padova, Italy}

\vspace{0.3cm}
\textit{\small ${}^e$ Department of Applied Mathematics and Theoretical Physics,\\ Centre for Mathematical Sciences, University of Cambridge,\\
Wilberforce Road, Cambridge, CB3 0WA, U.K.}

\end{center}

\vspace{3pt}

\abstract{We give a prescription for ${\cal N}=1$ supersymmetrization of any (four-dimensional) nonlinear electrodynamics theory with a Lagrangian density satisfying a convexity condition that we relate to semi-classical unitarity. We apply it to the one-parameter ModMax extension of Maxwell electrodynamics that preserves both electromagnetic duality and conformal invariance, and its  Born-Infeld-like generalization, proving that duality invariance is preserved. We also establish superconformal invariance of the superModMax theory by showing that its coupling to supergravity is super-Weyl invariant. The higher-derivative photino-field interactions that appear in any supersymmetric nonlinear electrodynamics theory are removed by an invertible nonlinear superfield redefinition.}

\bigskip

\thispagestyle{empty}


\newpage

\setcounter{footnote}{0}

\tableofcontents

\newpage

\section{Introduction}

In earlier work  we showed that the source-free Maxwell electrodynamics in a four-dimensional Minkowski spacetime
has a unique interacting one-parameter extension preserving both conformal invariance and electromagnetic duality invariance. The Lagrangian density is \cite{Bandos:2020jsw}
\begin{equation}\label{final-Lag11*}
{\cal L}_\gamma = (\cosh\gamma) S + (\sinh\gamma) \sqrt{S^2+P^2}\, ,
\end{equation}
where $\gamma$ is the parameter and $(S,P)$ are the Lorentz invariants quadratic in the
components of the two-form field strength $F=dA$ for a one-form potential $A$. In Minkowski coordinates $\{x^m; m=0,1,2,3\}$,
\be\label{SP=+}
S=- \frac 14 F_{mn}F^{mn}, \qquad P=-\frac 14 F_{mn}\tilde F^{mn}\,, \qquad \left(\tilde F^{mn}=\frac12 \varepsilon^{mnpq}F_{pq}\right)\,.
\ee
For $\gamma=0$ this Lagrangian density defines the source-free Maxwell theory, but for $\gamma>0$ it defines the interacting ``ModMax'' theory of \cite{Bandos:2020jsw}. As also explained in \cite{Bandos:2020jsw}, $\gamma<0$ can be excluded because it allows superluminal propagation of certain small-amplitude waves on any constant uniform electromagnetic background with non-zero $S^2+P^2$, whereas all small-amplitude waves are lightlike or subluminal for $\gamma>0$. This is one feature of the birefringence properties of ModMax electrodynamics, which are in accord with earlier results of
\cite{Denisov:2017qou} for conformal (but not necessarily duality) invariant nonlinear electrodynamics theories, or NEDs.  Other properties of ${\cal L}_\gamma$ have been discussed in  \cite{Kosyakov:2020wxv}.

The non-analyticity of ${\cal L}_\gamma$ at $S^2+P^2=0$ means that the Euler-Lagrange (EL) equations are ill-defined
for Lagrangian field configurations with $S=P=0$. However, the Hamiltonian field equations are analytic at the corresponding Hamiltonian field configurations; this is possible because the Legendre transform that takes ${\cal L}_\gamma$ to the Hamiltonian density ${\cal H}_\gamma$ maps configurations with $S=P=0$ to the boundary of the domain in which ${\cal H}_\gamma$ is convex \cite{Bandos:2020jsw}.
Exact solutions of the Hamiltonian field equations with $S=P=0$ include the vacuum and plane-waves \cite{Bandos:2020jsw}, and a class of topologically non-trivial configurations with knotted electromagnetic fields \cite{Dassy:2021ulu}.

It was also shown in \cite{Bandos:2020jsw} that ModMax electrodynamics is the weak-field limit
of a one-parameter duality-invariant generalization of Born-Infeld (BI) electrodynamics, although only the Hamiltonian density
of this BI-like theory was found there. The corresponding Lagrangian density is \cite{Bandos:2020hgy}
\begin{equation}\label{BIgammaL*}
  {\cal L}_{(\gamma BI)} = T - \sqrt{T^2 -2T\left[(\cosh\gamma)S +  (\sinh\gamma)\sqrt{S^2 +P^2}\right] - P^2}\, ,
\end{equation}
where $T$ is the BI constant with dimensions of energy density; for $\gamma=0$ we recover the BI theory for which $T$ can be interpreted (in a string-theory context) as the
D3-brane tension. The Lagrangian density of (\ref{final-Lag11*}) is recovered in the $T\to \infty$ limit, which is equivalent to a weak-field limit.

We should remark here that the coupling of generic nonlinear four-dimensional conformal electrodynamics theories to gravity has
been investigated in \cite{Denisova:2019lgr}, where it was shown that a certain condition on black hole charges restricts
the form of any conformal electrodynamics Lagrangian to a particular one-parameter extension of the Maxwell case. As we observed in
\cite{Bandos:2020hgy}, this class of conformal electrodynamics includes (after a constant rescaling of the gauge potential $A_m$)
the conformal {\it and} duality-invariant ModMax theory. More recently, investigations similar to those of \cite{Denisova:2019lgr} but
specific to ModMax have been carried out and further extended in \cite{Flores-Alfonso:2020euz,Bordo:2020aoz,Flores-Alfonso:2020nnd,Amirabi:2020mzv} (see also \cite{Neves:2021tbt,Bokulic:2021dtz,Mazharimousavi:2021jmn}).

Here we construct, using superfield methods,  the minimal (${\cal N}=1$) supersymmetric extension of both
the generalized BI theory described by (\ref{BIgammaL*}) and its weak-field limit (ModMax for $\gamma>0$). We do this by means of a general prescription that starts with any bosonic nonlinear electrodynamics theory for which the Lagrangian density is a strictly convex function of the electric field; as we show, this condition is required to ensure the absence of superluminal propagation of small-amplitude waves in a constant electromagnetic background. Given such a Lagrangian density as a function of $(S,P)$, this function is used to construct a single full superspace integral that yields a supersymmetrization of the initial bosonic theory, in the sense that the initial Lagrangian density is recovered by bosonic truncation.
We also extend this prescription to allow for spontaneously broken supersymmetry, although the intial nonlinear electrodynamics theory is then recovered only as the bosonic truncation of an effective field theory (in which the photino is now a goldstino) at energies much less than that of the supersymmetry breaking scale.

The electromagnetic duality invariance of ModMax and its BI-like generalization suggests that their supersymmetric extensions should have the same property. We show that the supersymmetric extensions provided by our general prescription indeed satisfy the previously established general conditions for duality invariance of super-NEDs \cite{Kuzenko:2000tg,Kuzenko:2000uh,Ivanov:2013ppa}.  The conformal invariance of ModMax similarly suggests that superModMax should be superconformal invariant and this is indeed the case.  Our proof  is indirect because
we first couple superModMax to supergravity and then show that the coupling is super-Weyl invariant, which implies that superModMax couples only to the fields of conformal supergravity. As superconformal invariance is the residual symmetry that results from imposing flat-superspace constraints on a super-Weyl and superdiffeomorphism theory, this establishes the superconformal invariance of superModMax. It should be appreciated that this was not guaranteed {\it a priori} because two supersymmetric extensions of a generic NED may differ by purely fermionic terms. We suspect, but do not prove, that the superModMax theory constructed here is the unique extension of the superMaxwell theory that is both superconformal invariant and duality invariant\footnote{ We would also expect supersymmetric Bialynicki-Birula electrodynamics \cite{Bialynicki-Birula:1984daz} to be both superconformal and duality invariant, but this would not be an `extension' of superMaxwell electrodynamics.}.

A general feature of supersymmetric nonlinear electrodynamics is the appearance of higher-derivative terms, but because these terms are nilpotent it is expected that they can be eliminated from the field equations (see e.g. \cite{Kuzenko:2005wh,Farakos:2013fne,Fujimori:2017kyi}). Here we show that all higher-derivatives can also be eliminated from the action by a non-linear superfield redefinition described in \cite{Cribiori:2018dlc}; this brings the Lagrangian density to the form
\begin{equation}\label{VAform}
{\cal L} = -{\cal L}_{VA}\,  \tilde{\cal L}_{\rm bos}
\end{equation}
where ${\cal L}_{VA}$ is the Volkov-Akulov (VA) Lagrangian density \cite{Volkov:1972jx,Volkov:1973ix}, i.e. the determinant of a goldstino-dependent generalization of the Minkowski vierbein, and $\tilde{\cal L}_{\rm bos}$ is the modification of the bosonic Lagrangian density obtained by using the associated goldstino-dependent metric in place of the Minkowski metric. This general result applies, in particular to superModMax, its superMaxwell limit and its BI-like generalization.
The appearance of the VA Lagrangian density as a factor in \eqref{VAform} does not imply spontaneous supersymmetry breaking unless
$\tilde{\cal L}_{\rm bos}$ has a non-zero vacuum value, in which case the vacuum energy will be positive and supersymmetry is spontaneously broken, with the photino field as the goldstino.
We shall discuss this VA formalism, along with its application to superModMax, towards the end of this paper.

We begin with a reminder of some basic facts about nonlinear supersymmetric electrodynamics, and a presentation of our general supersymmetrization prescription. We then show, in agreement with \cite{Shabad:2009ky}, that this convexity condition is a physical requirement that any nonlinear electrodynamics theory must satisfy if superluminal propagation is to be avoided in perturbations about a constant background electromagnetic field. These general results are then applied to ModMax and its BI-like extension.

\section{Super-electrodynamics preliminaries}\label{SNED}

All nonlinear $\mathcal N=1$ four-dimensional (4D) supersymmetric extensions of Maxwell electrodynamics are based on the same off-shell supermultiplet: the ``Maxwell supermultiplet''. For this reason it will be useful to begin with a rapid review of the superfield construction of supersymmetric electrodynamics in flat superspace. This will also serve to introduce conventions, which are mostly those of \cite{Wess:1992cp} and \cite{Buchbinder:1995uq}; details can be found in the Appendix.

The gauge invariant fields of the Maxwell supermultiplet appear in an anticommuting Weyl spinor chiral superfield $W_\alpha$. The chirality and ``superfield-Bianchi'' identities that it satisfies are
\be\label{bianchi}
\bar{\mathcal{D}}_{\dot\alpha} W_\alpha =0\, , \qquad
\mathcal{D}^\alpha W_\alpha-\mathcal{D}_{\dot\alpha} \bar{W}^{\dot{\alpha}} =0\, ,
\ee
where $(\mathcal{D}_\alpha, \bar{\mathcal{D}}_{\dot\alpha})$ are the Weyl-spinor supercovariant derivatives.
These constraints imply that
\be\label{WV}
W_\alpha = -\frac 14 \bar{\mathcal{D}}^2 \mathcal{D}_\alpha V\, , \qquad
\bar W_{\dot\alpha} = -\frac14 \mathcal{D}^2 \bar{\mathcal{D}}_{\dot\alpha} V\,, \qquad
\ee
where $V$ is an unconstrained real scalar ``prepotential'' superfield $V$, and we use the notation
\be
\mathcal{D}^2 = \mathcal{D}^\alpha \mathcal{D}_\alpha\, , \qquad
\bar{\mathcal{D}}^2=\bar{\mathcal{D}}_{\dot\alpha}\bar{\mathcal{D}}^{\dot\alpha} \, .
\ee
The superfield $V$, which has the 1-form potential $A_m$ as one of its components, is defined up to a superspace generalization of the abelian gauge transformation of $A_m$. The gauge-invariant independent component fields of $W_\alpha$ are given by
\begin{eqnarray}\label{Wcpts}
W_\alpha| &=& -i\lambda_\alpha \nonumber \\
(\mathcal{D}_{(\alpha} W_{\beta)})| &=&  -(i/2)F_{mn} (\sigma^{mn})_{\alpha\beta} \nonumber \\
(\mathcal{D}^\alpha W_\alpha)| &=& -2D\, ,
\end{eqnarray}
where the vertical bar indicates that we set to zero the anticommuting spinor superspace coordinates  ($\theta^\alpha, \bar\theta^{\dot\alpha}$).
The complex Weyl spinor field $\lambda_\alpha$ is the photino field (its complex conjugate
is $\bar\lambda_{\dot\alpha} = -i\bar W_{\dot\alpha}|$) and $D$ is the auxiliary field. The equivalent expansion in powers of anticommuting spinor coordinates is
\be
\label{Wexp}
 W_\alpha =e^{\ii\theta\sigma^l\bar\theta\partial_l}\left(-\ii\lambda_\alpha +\theta_\alpha D(x)- \frac \ii 2 F_{mn}(\sigma^{mn}){}_\alpha{}^\beta\theta_\beta
+\theta^2 (\sigma^{m}_{\alpha\dot\beta}\,
\partial_m\bar{\lambda}{}^{\dot\beta}) \right)\, .
\ee
The infinitesimal supersymmetry transformations of the
component fields, with constant Weyl spinor parameter $\epsilon$, are
\bea
\delta_\epsilon A_m &=& i\epsilon\sigma_m\bar\lambda-i\lambda\sigma_m\bar\epsilon \, , \qquad \delta_\epsilon D = -\partial_m(\epsilon\sigma^m\bar\lambda+\lambda\sigma^m\bar\epsilon), \nonumber \\
\delta_\epsilon\lambda^\alpha &=& i \epsilon^\alpha  D - \frac 1 2 (\epsilon\,\sigma^{mn})^\alpha F_{mn}\, .
\eea

The nilpotent chiral scalar superfield $W^2=W^\alpha W_\alpha$ will play an important role in what follows; its components are defined by
\begin{eqnarray}
W^2| &=& -\lambda^2 \qquad (\lambda^2 = \lambda^\alpha\lambda_\alpha) \label{lam2} \\
 \frac 1 {\sqrt{2}}(\mathcal{D}_\alpha W^2)| &=& \chi_\alpha =\sqrt{2}\,\left(\frac12 \, F_{mn}(\sigma^{mn}){}_{\alpha}{}^\beta\lambda_\beta-\ii\lambda_\alpha D \right),\label{chi} \\
-\frac 14(\mathcal{D}^2 W^2)| &=&  F = 2\left( S+\ii P + \frac12 D^2 - \ii \lambda\sigma^m \partial_m\bar\lambda\right)\label{F=2} \, ,
\end{eqnarray}
where $S$ and $P$ are the Lorentz scalars defined in (\ref{SP=+}).

What we now need is a superspace Lagrangian density that generalises a Lagrangian density
${\mathcal L}(S,P)$ of a generic (non-linear) electrodynamics theory. A straightforward way to achieve this is to consider the following choice\footnote{This differs (in form) from, but is related to, the structure of supersymmetric non-linear electrodynamics Lagrangians considered previously; see e.g. \cite{Deser:1980ck,Cecotti:1986gb,Bagger:1996wp,Fujimori:2017kyi}.}
\be\label{Lchoice}
{\cal L} =  \int\! d^2\theta d^2\bar\theta \frac{16W^2 \bar W^2}{\mathcal{D}^2 W^2 \bar{\mathcal{D}}^2 \bar W^2} \,
L(\mathbb S,\mathbb P, \mathbb D)\, ,
\ee
where
\begin{eqnarray}\label{SPSF}
{\mathbb S}&=& -\frac 1{16}(\mathcal{D}^2W^2+\bar{\mathcal{D}}^2\bar W^2)\, ,  \quad
{\mathbb P}=\frac i{16}(\mathcal{D}^2W^2-\bar{\mathcal{D}}^2\bar W^2)\, , \quad \mathbb D=\frac 12 \mathcal{D}_\alpha W^\alpha\, .
\end{eqnarray}
The respective leading components of these three superfields are, for zero fermion fields,  $(S,P,D)$.
To  evaluate the Berezin integral of (\ref{Lchoice}) we may use the identity
\be \label{Bint-id}
\int d^2\theta d^2 \bar\theta \equiv \frac{1}{32}  \left\{ \mathcal{D}^2,\bar{\mathcal{D}}^2\right\}\, .
\ee
When this superspace differential operator acts on the integrand of (\ref{Lchoice}), all terms will be at least quadratic in the photino field except those for which
$\mathcal{D}^2$ acts on $W^2$ and $\bar{\mathcal{D}}^2$ acts on $\bar W^2$, which cancels the $\mathcal{D}^2 W^2 \bar{\mathcal{D}}^2 \bar W^2$ denominator.
The final result, omitting total derivative terms, therefore takes the form
\be
{\mathcal L} = L\left(S+ \tfrac12 D^2,P,D\right) + {\cal O}(\lambda^2)\, .
\ee
At $\lambda=0$ the field equation for $D$ is
\be\label{Deom}
L_S D + L_D = 0 \, , \qquad (L_X \equiv \partial L/\partial X).
\ee

We would like this to have a unique solution for $D$. Given that a solution exists, 
a sufficient condition for it to be unique is strict convexity of $L$ as a function of $D$, which requires 
\be\label{convexityD}
L_S + L_{DD} + D^2L_{SS} + 2D L_{SD} > 0\, .
\ee
If $D=0$ is the unique solution of \eqref{Deom} then 
\be
{\mathcal L}_{\lambda=0} = L(S,P, 0) \, .
\ee
However,  there are many three-variable bosonic truncations of  $L(\mathbb S,\mathbb P, \mathbb D)$ that lead to the same two-variable function $L(S,P,0)$; this illustrates the well-known fact that two supersymmetric Lagrangians  which reduce to the same bosonic one may differ by fermionic terms \cite{Cecotti:1986gb,Bagger:1996wp}.

For this reason we shall now focus on a special class of three-variable functions 
$L(\mathbb S,\mathbb P, \mathbb D)$. It is a class for which the convexity condition 
\eqref{convexityD}, which is sufficient for uniqueness of a solution of the $D$-field equation, 
is also necessary for reasons that will be explained in section \ref{sec:convexity}.

\subsection{Minimal supersymmetrization}

We shall consider supersymmetric nonlinear electrodynamics theories that can be found from the choice
\be\label{Simplechoice1}
L(\mathbb S,\mathbb P, \mathbb D) = \mathcal{L}^{\rm bos}(\mathbb S,\mathbb P) - \xi {\mathbb D} \, ,
\ee
where the function ${\cal L}^{\rm bos}$ is the Lagrangian density of some bosonic theory of interest but with
$(S,P)$ replaced with $(\mathbb S,\mathbb P)$. The constant $\xi$ has the same dimensions as the Fayet-Iliopoulos (FI) constant
but the super-invariant generated by $\xi{\mathbb D}$, when used in \eqref{Lchoice}, differs from the standard FI term discussed in the super-NED context in \cite{Kuzenko:2013gr}. Whereas
the component FI term is $\xi D$, the component version of the super-invariant generated by $\xi {\mathbb D}$, in the construction
based on \eqref{Lchoice}, includes additional fermionic interaction terms \cite{Cribiori:2017laj,Kuzenko:2018jlz}.

Omitting fermions, the full component Lagrangian density resulting from the choice \eqref{Simplechoice1} is
\be\label{returnto}
{\mathcal L}_{\lambda=0} = {\cal L}^{\rm bos}(S+ \tfrac12 D^2,P) - \xi D\, ,
\ee
and the $D$ field equation \eqref{Deom} simplifies to
\be\label{Deom2}
{\mathcal L}^{\rm bos}_SD = \xi + {\cal O}(\lambda^2)\, .
\ee
The convexity condition \eqref{convexityD} also simplifies,  to
\be
{\cal L}^{\rm bos}_S + D^2 {\mathcal L}^{\rm bos}_{SS} >0\, ,
\ee
which is satisfied for all $D$ provided that
\be\label{convexity2}
{\mathcal L}^{\rm bos}_S >0 \, , \qquad {\mathcal L}^{\rm bos}_{SS}\ge 0\, .
\ee
The necessity of these conditions will be explained in section \ref{sec:convexity}. From this fact and \eqref{Deom2} we see that 
$D|_{\lambda=0}=0$  (and hence supersymmetry is unbroken) iff $\xi=0$. For this case we have
\be\label{simplechoice}
{\cal L} :=  \int\! d^2\theta d^2\bar\theta\, \frac{16W^2 \bar W^2}{\mathcal{D}^2 W^2 \bar{\mathcal{D}}^2 \bar W^2} \,
{\mathcal L}^{\rm bos}(\mathbb S,\mathbb P)  = {\mathcal L}^{\rm bos}(S,P) + {\cal O}(\lambda^2)\, ,
\ee
which  provides a simple prescription for the supersymmetrization of any bosonic nonlinear electrodynamics theory with a Lagrangian density satisfying \eqref{convexity2}.

This could be viewed as a ``minimal'' prescription; we shall use it extensively in the remainder of this paper.
As an illustration of its use, we apply it to Maxwell electrodynamics, for which ${\cal L}^{\rm bos}=S$. The corresponding
superspace action is
\bea \label{Maxchoice}
{\cal L}_{SM} &=& -  \int\! d^2\theta d^2\bar\theta\,  \frac{W^2 \bar W^2}{\mathcal{D}^2 W^2 \bar{\mathcal{D}}^2 \bar W^2}\,
{\mathbb S} \nonumber \\
&=& -\int\! d^2\theta d^2\bar\theta \left\{ W^2 \frac{\bar W^2}{\mathcal{D}^2 W^2}
+ \bar W^2 \frac{ W^2}{\bar{\mathcal{D}}^2\bar W^2} \right\} \nonumber\\
&=&\frac 14 \int \! d^2\theta\,  W^2 + c.c.
\eea
which is the standard superfield form of the super-Maxwell action. Evaluation of the Berezin integral yields the standard free-field component
result:
\begin{equation}\label{sMax}
{\cal L}_{SM} = \frac14 F + c.c. = S -\frac{i}{2} \lambda\sigma^m\,  \overleftrightarrow{\partial_m}\,  \bar\lambda + \frac12 D^2\, .
\end{equation}

\subsubsection{Spontaneously broken supersymmetry}\label{FI}

More generally, allowing for arbitrary $\xi$, the elimination of $D$ from \eqref{returnto} is equivalent to taking the Legendre transform of ${\mathcal L}_{\lambda=0}$ viewed as a (strictly convex) function of $D$. The result is
\be
-{\mathcal L}_{\lambda=0}(\xi) := \sup_D \left\{\xi D -
{\mathcal L}_{\lambda=0}(D)\right\}\, .
\ee
By construction, the left hand side is (i) a strictly convex function of $\xi$ with a unique global minimum,
which must be zero for $\xi=0$, because supersymmetry is unbroken in this case, and (ii) strictly positive away from this global minimum, which implies that supersymmetry is spontaneously broken for $\xi\ne0$; this requires $D\ne0$ and hence, on dimensional grounds, $D\propto \xi$.  These results would be a consequence of supersymmetry alone if we could assume the absence of negative energy solutions of the field equations (the classical analog of ghosts), which suggests that the convexity conditions (\ref{convexity2}) are necessary for the absence of ghosts; we elaborate on this in the following section.

Another implication of $\xi\ne0$ or, equivalently, $D\ne0$, is that ${\mathcal L}_{\lambda=0}$ is now the Lagrangian density of the initial bosonic theory that we are attempting to supersymmetrize only at energies much less than $\xi^2$; i.e
\be
{\mathcal L}_{\lambda=0} = {\mathcal L}^{\rm bos}(S,P) + {\cal O}(\xi^2) \, .
\ee
This is not the full effective Lagrangian since the photino, which is now a goldstino, remains massless. However, the photino/goldstino will become the spin-$\frac 12$ polarisation states of a gravitino with mass proportional to $\xi^2$ after coupling to supergravity.

\section{Physics of convexity}\label{sec:convexity}

For any Lagrangian density ${\cal L}(S,P)$, we have
\begin{equation}
\frac{\partial {\mathcal L}}{\partial {\bf E}} = {\cal L}_S {\bf E} + {\mathcal L}_P {\bf B} \, ,
\end{equation}
and the matrix of second-derivatives with respect to the electric field components $\{E_i; \,i=1,2,3\}$ is
\begin{equation}\label{Hess}
H_{ij} = {\mathcal L}_S \delta_{ij} + {\mathcal L}_{SS} E_iE_j +
{\mathcal L}_{SP} (E_iB_j+ E_jB_i)+ {\mathcal L}_{PP} B_iB_j\, .
\end{equation}
This is the Hessian, which must be strictly positive for strict convexity of ${\mathcal L}$; i.e. its eigenvalues must be positive.
These eigenvalues are
\be\label{evals}
{\mathcal L}_S, \qquad {\mathcal L}_{S} + X- \sqrt{X^2-Y}\, , \qquad {\mathcal L}_S +X + \sqrt{X^2-Y}\, ,
\ee
where
\bea\label{XYexpressions}
X &=&\frac12 \left({\mathcal L}_{SS} |{\bf E}|^2 + 2P {\mathcal L}_{SP} + {\mathcal L}_{PP} |{\bf B}|^2\right)\, , \nonumber \\
Y &=& \left[{\mathcal L}_{SS} {\mathcal L}_{PP} - {\mathcal L}_{SP}^2\right]|{\bf E}\times {\bf B}|^2 \, .
\eea

Convexity is an important consideration as it is related to the absence, in a semi-classical context,  of both ghosts (negative energy) and
tachyons (imaginary mass); i.e, to semi-classical unitarity. The link to energy and the issue of whether it is positive comes from consideration of the
Hamiltonian density defined by the Legendre transform of ${\mathcal L}$; this is
\be
{\mathcal H}({\bf D}, {\bf B}) = \sup_{{\bf E}} \left[ {\bf D}\cdot{\bf E} - {\mathcal L}({\bf E},{\bf B}) \right]\, ,
\ee
where ${\bf D}$ is the ``electric displacement'' field, which is canonically conjugate to the electric field ${\bf E}$. By construction, this Hamiltonian density is a convex function of ${\bf D}$, which implies that any local minimum (with respect to variations of ${\bf D}$ for given ${\bf B}$) is a global minimum, which is unique if  $\mathcal H$ is strictly convex; this is the vacuum, and no field configuration has lower energy. However, the initial Lagrangian density ${\mathcal L}({\bf E},{\bf B})$ will be recovered from ${\mathcal H}({\bf D}, {\bf B})$ by a Legendre transform with respect to ${\bf D}$ only if it is a convex function of ${\bf E}$. If it is not convex then, typically, the initial function will be recovered but with a restriction on its domain (the domain in which, for given ${\bf B}$, it is a convex function of ${\bf E}$). Outside this domain, one may find a formal ``Hamiltonian density'' from some extremal value of ${\bf D}\cdot{\bf E} - {\mathcal L}$ that is {\it not} a maximum (with respect to variations of ${\bf E}$) but it will generically have values that are lower than the vacuum energy. The convexity condition on ${\mathcal L}$ is therefore a standard condition for semi-classical unitarity, and in practice one requires strict convexity because
otherwise  ${\bf D}$ is not uniquely defined\footnote{Despite this, the Legendre transform remains involutive, as shown in \cite{Bandos:2020hgy} for the case of Bialynicki-Birula electrodynamics where the non-zero Hamiltonian density is the Legendre transform of a Lagrangian density that is identically zero!} for given $({\bf E},{\bf B})$ by maximisation of
${\bf D}\cdot {\bf E} -\mathcal L$.

For Lorentz invariant theories, the conditions for strict convexity are Lorentz invariant because the Legendre transform that provides an involutive one-to-one map between convex Hamiltonian and Lagrangian densities also preserves Lorentz invariance, even though this symmetry is not manifest in the Hamiltonian formulation; this is because the Legendre transform is effected, essentially, by a process of algebraic elimination of an `auxiliary' vector field.  As we show below, this allows a simple determination of the Lagrangian convexity conditions by an appropriate choice of inertial frame. We then show how the same conditions arise by requiring the absence of superluminal small-amplitude waves in a constant uniform background magnetic field. We then explore  this connection between convexity and causality for general background fields in the context of conformal electrodynamics, with results that generalize our earlier ModMax results\cite{Bandos:2020jsw} and  complement those of \cite{Shabad:2009ky,Denisov:2017qou}.

From \eqref{evals} we see that  strict convexity always requires ${\mathcal L}_S>0$, and this allows us to introduce the notation
\be
\ell_{SS} = \frac{{\mathcal L}_{SS}}{{\mathcal L}_S} \, , \qquad \ell_{SP} = \frac{{\mathcal L}_{SP}}{{\mathcal L}_S} \, , \qquad
\ell_{PP} = \frac{{\mathcal L}_{PP}}{{\mathcal L}_S} \, .
\ee

Using the identities
\be
|{\bf E}|^2 \equiv  |{\bf B}|^2 +2S \, , \qquad |{\bf E} \times {\bf B}|^2 \equiv  |{\bf B}|^4 +2S |{\bf B}|^2 - P^2\, ,
\ee
we may rewrite the expressions for $(X,Y)$ in the form
\bea
X &=& {\mathcal L}_S \left[\Xi |{\bf B}|^2  + S\ell_{SS} + P\ell_{SP} \right] \, , \nonumber \\
Y &=& {\mathcal L}^2_S\,   \left[ |{\bf B}|^4  + 2S |{\bf B}|^2 - P^2\right]\Gamma \, ,
\eea
where
\be\label{GammaXi}
\Xi =  \frac12(\ell_{SS} +\ell_{PP} )\, , \qquad \Gamma = \ell_{SS}\ell_{PP} - \ell_{SP}^2 \, .
\ee
The point of this is that a Lorentz boost will not change $(S,P)$ but we may boost to a frame in which $|{\bf B}|$ is arbitrarily large, and this yields
\be\label{preZ}
X \pm \sqrt{X^2-Y} = Z_\pm |{\bf B}|^2 + {\mathcal L}_S\left[\Upsilon \pm \frac{\Xi\Upsilon - S\Gamma}{\sqrt{\Xi^2-\Gamma}} \right] + {\mathcal O}(1/|{\bf B}|^2)\, , 
\ee
where
\be
Z_\pm = {\mathcal L}_S \left\{ \Xi \pm \sqrt{\Xi^2 - \Gamma} \right\}\, ,  \qquad \Upsilon = S\ell_{SS} + P\ell_{SP} \, .
\ee
The eigenvalues of the Hessian matrix in this frame are therefore
\be\label{largeB}
{\mathcal L}_S \, , \qquad  Z_- |{\bf B}|^2 +  {\mathcal O}(1) \, , \qquad  Z_+ |{\bf B}|^2 +  {\mathcal O}(1)\, .
\ee
We conclude that the Hessian  will not be positive if  either $Z_+$ or $Z_-$ is negative, so  we require $Z_\pm\ge0$ and this will be satisfied if
$Z_-\ge0$, which is equivalent to
\be\label{XiGam}
\Xi  \ge \sqrt{\Xi^2 -\Gamma}\, .
\ee
This requires $\Gamma\ge0$ and implies $\Xi\ge0$. We have equality  (equivalently $Z_-=0$) iff $\Gamma=0$. To determine whether this is allowed by convexity
we need to consider the ${\mathcal O}(1)$ terms in \eqref{largeB}, but we need this only for $\Gamma=0$:
\begin{itemize}

\item $\Gamma=0$. In this case the eigenvalues are
\be
{\mathcal L}_S, \qquad {\mathcal L}_S\, , \qquad {\mathcal L}_S \left[ 1+ 2 \left(\Xi |{\bf B}|^2  + \Upsilon \right)\right] + {\mathcal O}(1/|{\bf B}|^2) \, .
\ee
By going to a Lorentz frame in which $|{\bf B}|$ is sufficiently large, we see that all eigenvalues are positive (for positive ${\mathcal L}_S$) when
$\Xi>0$, while  $\Xi=0$ implies $\Upsilon=0$ when $\Gamma=0$, in which case the third eigenvalue  equals the other two.
\end{itemize}
This confirms that equality in \eqref{XiGam} is permitted by strict convexity.  We thus deduce, by combining the conditions $\Xi\ge0$ and $\Gamma\ge0$ with ${\mathcal L}_S>0$, that strict convexity of $\mathcal L$ requires
\be\label{concon1}
{\mathcal L}_S >0 \, , \qquad
{\mathcal L}_{SS} \ge 0 \, , \qquad {\mathcal L}_{PP} \ge 0 \, ,
\ee
and
\be\label{concon2}
{\mathcal L}^2_{SP} \le {\mathcal L}_{SS} {\mathcal L}_{PP}\, .
\ee
Notice that  the conditions \eqref{concon1} include those of \eqref{convexity2}, which we deduced by requiring convexity of the bosonic truncation of the superspace Lagrangian density of super-NEDs as a function of the auxiliary field $D$. In fact, the conditions \eqref{concon1} and \eqref{concon2} reduce to those of \eqref{convexity2} for the special case in which ${\mathcal L}_P\equiv 0$; this is no coincidence because the $D$-dependence of ${\mathcal L}(S +\tfrac12 D^2)$ is the same as its dependence on any Cartesian component of the electric field.

We now turn to the topic of small-amplitude plane waves in a constant uniform background electromagnetic field strength $F_{mn}$. Linearization of the  EL equations for generic Lagrangian
density ${\mathcal L}(S,P)$,  about such a background, yields a two-polarization wave-equation for small-amplitude disturbances.  The dispersion relations for the two polarization modes are generally different  and they take the form\footnote{This corrects a sign error in \cite{Bandos:2020jsw}.}
\begin{equation}\label{disp}
k^2 = - G^2 \lambda_\pm\, ,
\end{equation}
where $k$ is wave 4-vector and $G^2$ is the norm of the 4-vector $G_m =F_{mn}k^n$. The coefficients $\lambda_\pm$ are called the
``birefringence indices'', and there will be superluminal wave-propagation if either
$\lambda_+$ or $\lambda_-$ is negative. We shall therefore require $\lambda_\pm \ge0$. A calculation shows that  \cite{Bialynicki-Birula:1984daz}
\begin{equation}\label{biref}
\lambda_\pm = \frac12  [P^2\Gamma + 2(S\ell_{PP} - P\ell_{SP}) -1]^{-1}\left[ (2S\Gamma - \ell_{SS}-\ell_{PP}) \mp \sqrt{\Delta}\right\}\, ,
\end{equation}
where $\Gamma$ is given in \eqref{GammaXi}, and
\begin{equation}
\Delta = (\ell_{SS} - \ell_{PP} -2S\Gamma)^2 + 4(\ell_{SP} - P\Gamma)^2\, .
\end{equation}

We can make contact with the  constraints imposed by convexity of ${\mathcal L}$ as a function of the electric field by choosing a purely magnetic background.  In this case, we have
\be
\Delta|_{|{\bf E}|=0} = \left(2\Xi + \Gamma|{\bf B}|^2\right)^2 -4A \Gamma\, , \qquad A :=  1+ \ell_{PP} |{\bf B}|^2\, , 
\ee
where $\Xi$ is the expression given in \eqref{GammaXi}. This yields
\be
\lambda_\pm |_{|{\bf E}|=0} =  \frac{1}{2A} \left[ 2\Xi +  \Gamma |{\bf B}|^2 \pm \sqrt{(2\Xi +  \Gamma|{\bf B}|^2)^2 - 4A\Gamma} \right]\,  .  
\ee
In order to ensure that the indices are non-negative for any value of $|{\bf B}|^2$ we require
\be
\ell_{PP}\ge0 \, , \qquad \Xi\ge0 \, , \qquad \Gamma\ge 0 \qquad (\Rightarrow \ \ell_{SS}\ge0) \, .
\ee
Taking into account the  condition ${\mathcal L}_S>0$, these are precisely the convexity constraints of \eqref{concon1} and \eqref{concon2}.

The relation between convexity and the absence of superluminal small-amplitude waves is more complicated for other backgrounds, so let us
again consider the special class of theories for which $\Gamma=0$:
\begin{itemize}

\item $\Gamma=0$. In this case $\sqrt{\Delta} = 2|\Xi|$. On the assumption that $\Xi\ge0$, we have
\be\label{gamzero2}
\lambda_- = 0\, , \qquad \lambda_+ = \frac{2\Xi}{1- 2S\ell_{PP} + 2P\ell_{SP}}\, .
\ee
The opposite assumption, that $\Xi\le0$, gives the same result but with the roles of $\lambda_\pm$ interchanged, so we can use the above formula
for $\lambda_+$ as the formula for the not-necessarily-zero index for either sign of $\Xi$. However, consideration of purely magnetic backgrounds shows that
$\lambda_\pm\ge0$ requires  both $\Xi\ge0$ and $\ell_{PP}\ge0$, and hence $\ell_{SS}\ge0$.  This confirms our earlier general result for purely magnetic backgrounds, but for more general backgrounds an additional constraint may be needed (depending on the theory) to ensure positivity  of the denominator of the expression for $\lambda_+$.
\end{itemize}

A particular subclass of theories with $\Gamma=0$, to which we now turn our attention,  is conformal electrodynamics.

\subsection{Conformal electrodynamics}

The absence of any dimensionful constants in conformal theories is expressed by the Euler relation
\be
S{\mathcal L}_S + P{\mathcal L}_P = \mathcal L\, .
\ee
This implies
\be\label{diffSP}
S{\mathcal L}_{SP} + P{\mathcal L}_{PP} =0\, , \qquad S{\mathcal L}_{SS} + P{\mathcal L}_{SP}=0\, , \qquad S^2{\mathcal L}_{SS} -P^2{\mathcal L}_{PP} =0\, ,
\ee
and hence
\be\label{integrability}
{\mathcal L}_{SS} {\mathcal L}_{PP} -{\mathcal L}_{SP}^2 =0\, .
\ee
We may use these relations in \eqref{XYexpressions} to deduce that
\be
X = {\mathcal L}_S\,  \Xi \,  |{\bf B}|^2 \, , \qquad Y=0\, .
\ee
The Hessian matrix eigenvalues for conformal electrodynamics theories are therefore
\be
{\mathcal L}_S, \qquad {\mathcal L}_{S}\, , \qquad {\mathcal L}_S\,  \left[ 1 + \Xi  |{\bf B}|^2\right] \, .
\ee{}
We thus confirm that strict convexity requires $\Xi\ge 0$ in addition to ${\mathcal L}_S>0$. Since $\Gamma=0$,
this also requires both $\ell_{SS}\ge 0$ and $\ell_{PP}\ge 0$. We thus recover the convexity inequalities \eqref{concon1}, while
the inequality \eqref{concon2} is saturated.

We now turn to the birefringence properties of conformal electrodynamics.
The relations \eqref{diffSP} may be used to rewrite the expressions of \eqref{gamzero2} for $\lambda_\pm$ as
\be\label{lpm}
\lambda_-=0\, , \qquad \lambda_+ = \frac{ (S^2+P^2)\ell_{PP}}{S^2  - 2S(S^2+P^2)\ell_{PP}} \, .
\ee
In a purely magnetic background we require $\ell_{PP} \ge 0$ to ensure $\lambda_+\ge0$ for any choice of $|{\bf B}|$, and this implies that
$\ell_{SS}\ge0$. Given that ${\mathcal L}_S>0$ (which is certainly required for positive energy in the ModMax subcase) we recover the
conditions required by convexity, as already noted for the general $\Gamma=0$ case. However, for a general (constant and uniform) background
we also require
\be\label{adcon}
S^2 {\mathcal L}_S> 2S(S^2+P^2){\mathcal L}_{PP}  \, ,
\ee
which is not obviously satisfied for $S>0$, e.g. a purely electric background.  Generically, \eqref{adcon} is a {\it stronger} constraint than ${\mathcal L}_S>0$, but not for the ModMax case of most interest here, as we shall now see.

It is instructive to consider ModMax electrodynamics within the slightly larger class of conformal electrodynamics
defined by the following Lagrangian density in which $(\alpha,\beta)$ are arbitrary constants:
\be\label{instruct}
\mathcal L = \alpha S + \beta \sqrt{S^2+P^2} \, .
\ee
This yields
\be\label{MMData}
{\mathcal L}_S = \alpha+ \beta \frac{S}{\sqrt{S^2+P^2}} \, , \qquad (S^2+P^2) {\mathcal L}_{PP} = \beta \frac{S^2}{\sqrt{S^2+P^2}} \, ,
\ee
We see that $\alpha >|\beta|$ is required for ${\mathcal L}_S>0$ (which is required for positive energy, as can be verified using steps spelled out in \cite{Bandos:2020jsw}).
The convexity condition ${\mathcal L}_{PP} \ge0$ then requires $\beta\ge 0$, so that positive energy and convexity imply $\alpha >\beta \ge0$.  These conditions are not obviously sufficient
for $\lambda_+\ge0$ because this requires
\be
0 < \, S^2 {\mathcal L}_S -2S(S^2+P^2) {\mathcal L}_{PP} = S^2\left[ \alpha - \beta \frac{S}{\sqrt{S^2+P^2}} \right]\, ,
\ee
where the equality follows from \eqref{MMData}; the resulting inequality is equivalent to ${\mathcal L}_S>0$. In this case, therefore, positive energy and convexity are jointly equivalent to the requirement of non-negative birefringence indices, for any choice of the constant uniform background for which $S^2+P^2\ne0$.

As a final point we observe that the physical restrictions $\alpha >\beta \ge0$ imposed on the model defined by
\eqref{instruct} are solved by setting
\be
(\alpha, \beta) = a(\cosh\gamma, \sinh\gamma) \,  \qquad a>0\, , \quad \gamma\ge 0\, ,
\ee
but ${\mathcal L}$ is then just a positive constant times the ModMax Lagrangian density, and hence has EL equations that are duality invariant.
In other words, the duality-invariant ModMax theory is essentially the only physical acceptable model with a Lagrangian density of the
form \eqref{instruct}.

\section{SuperModMax}

To apply the supersymmetrisation procedure described in the previous section to ModMax electrodynamics, we start from a superfield Lagrangian of the form (\ref{Lchoice}) with
\be\label{MMchoice}
L({\mathbb S},{\mathbb P}) = (\cosh\gamma) {\mathbb S}\,  + \sinh\gamma \sqrt{ {\mathbb S}^2 + {\mathbb P}^2}\, .
\ee
The resulting superfield Lagrangian density  is equivalent to
\be\label{SMM}
{\mathcal L}_{SMM}=\frac14 \left\{ \cosh\gamma \left[ \int\! d^2\theta\,  W^2 + c.c. \right] +
2\sinh\gamma \int\! d^2\theta d^2\bar\theta \frac{W^2 \bar W^2}{\sqrt{{\mathcal D}^2 W^2 \bar {\mathcal D}^2 \bar W^2}}\right\}\, .
\ee
The bosonic truncation of the complete component Lagrangian density (which will be discussed in section \ref{sec:cpts}) is
\be\label{cMM}
{\mathcal L}^{({\rm bos})}_{SMM}=\cosh\gamma \left(S+\tfrac12 D^2\right)+\sinh\gamma\,
\sqrt{\left(S+\tfrac12 D^2\right)^2+P^2}\,.
\ee
For $\gamma=0$ we recover  (\ref{sMax}). For $\gamma \ne 0$, the field equation for $D$ is equivalent to
\be\label{Deq}
D\left(\coth\gamma +  \frac{S+\tfrac12 D^2}{ \sqrt{\left(S+\tfrac12 D^2\right)^2+P^2}}\right)=0.
\ee
The unique solution is $D=0$, and we thus recover the Lagrangian density ${\mathcal L}_\gamma$ of (\ref{final-Lag11*}) by elimination of 
the auxiliary field $D$. 

For $\gamma>0$ this result was guaranteed by the strict convexity, as a function of $D$, of the expression of \eqref{cMM} for ${\mathcal L}^{({\rm bos})}_{SMM}$, but it is also true for $\gamma<0$ despite the non-convexity of the same expression in this case. This illustrates the fact that strict convexity as a function of $D$ is not necessary for uniqueness of the solution for $D$; it is only a sufficient condition. However, strict convexity {\it is} necessary in the context of an extension to include a supersymmetry-breaking FI-type term, for reasons explained in subsection \ref{FI}. In addition,  it should also be remembered that convexity as a function of $D$ follows (for the minimal super-NEDs considered here, which include ModMax) from convexity as a function of the electric field, which is needed for the reasons spelled out in section \ref{sec:convexity}, and this requires $\gamma>0$ for ModMax.

\subsection{Duality invariance}

The conditions required for electromagnetic duality invariance of generic nonlinear electrodynamics theories \cite{Gaillard:1981rj,Gibbons:1995cv,Gaillard:1997rt,Gaillard:1997zr} were generalized
by Kuzenko and Theisen to superfield formulations of $\mathcal N=1$ and $\mathcal N=2$ supersymmetric theories \cite{Kuzenko:2000tg,Kuzenko:2000uh} (see  \cite{Ivanov:2013ppa} for further developments); they also proposed a  perturbative scheme to compute duality invariant ${\mathcal N} = 2$ superconformal actions. For generic $\mathcal N=1$ theories described by an action $I[ W, \bar{ W}]$, the Kuzenko--Theisen duality-invariance condition is
\be\label{SDC}
\Im\int d^4x\,d\theta^2\left( W^\alpha W_\alpha+M^\alpha M_\alpha \right)=0\, , \qquad
M_\alpha\equiv-2i\frac{\delta I[W,\bar W]}{\delta W^\alpha} \, ,
\ee
where the super-Bianchi identity of \eqref{bianchi} should not be imposed on $W$ here (because duality transformations
act on field-strengths rather than potentials).

For the action with superModMax superfield Lagrangian density (\ref{SMM}),  we have
\be\label{MSMM}
M_\alpha= -iW_\alpha\left[\cosh\gamma - 2(\sinh\gamma)\,\bar {\mathcal D}^2 J\right]\, ,
\ee
with
\be
J = \bar W^2 \left[\frac 1{\sqrt{{\mathcal D}^2W^2\bar{\mathcal D}^2\bar W^2}}
-\frac 12 {\mathcal D}^2\left(\frac {W^2\bar{\mathcal D}^2\bar W^2}{({\mathcal D}^2W^2\bar{\mathcal D}^2\bar W^2)^{\frac 32}}\right)\right]\, .
\ee
This yields
\begin{eqnarray}
W^2 + M^2 &=& -(\sinh\gamma)^2 W^2 \left\{1 + \left[\bar {\mathcal D}^2 \left(\frac{\bar W^2}{\sqrt{{\mathcal D}^2 W^2 \bar{\mathcal D}^2\bar W^2}} \right)\right]^2\right\} \nonumber \\
&& +\  2(\sinh\gamma)(\cosh\gamma) W^2 \bar {\mathcal D}^2 \left(\frac{\bar W^2}{\sqrt{{\mathcal D}^2 W^2\bar{\mathcal D}^2 \bar W^2}}\right) \,.
\end{eqnarray}
Using this expression, and the fact that $W_\alpha W^2 \equiv 0$, one may verify that
the duality-invariance condition \eqref{SDC} is satisfied.

\subsection{Coupling to supergravity}

A generic supersymmetric non-linear electrodynamics with a Lagrangian density of the form \eqref{Lchoice}  can be coupled to supergravity as follows \cite{Kuzenko:2002vk,Kuzenko:2005wh}.
The supergravity fields are contained in the supervielbein:  $E_M{}^A$ in local superspace coordinates $z^M$, where $A= (a, \alpha, \dot\alpha)$ are vector and spinor indices
of $SL(2;\mathbb{C})$. The superspace integrand is the product of $E$, the Berezinian (superdeterminant) of this supervielbein, with a scalar constructed from the chiral
field-strength superfield $W$, its supercovariant derivatives, and their complex conjugates. The construction proceeds by direct analogy with \eqref{Lchoice} but the supercovariant derivatives are now
the spinor components of the covariant exterior derivative ${\bf{D}} = dz^M E_M{}^A {\mathfrak{D}}_A$; these have the property that
\be
\{\bar{\mathfrak D}_{\dot\alpha},\bar{\mathfrak D}_{\dot\beta}\}=-8R \, \bar M_{\dot\alpha\dot\beta}\, ,
\ee
where $\bar M_{\dot\alpha\dot\beta}$ are $SL(2,C)$ generators and $R$ is a scalar chiral superfield formed from the purely spinor components of the superspace curvature tensor  (see e.g. \cite{Wess:1992cp,Buchbinder:1995uq}). This leads to a generalization of the flat superspace chiral superfield $\bar{\mathcal D}^2\bar W^2$:
\be\label{defU}
U := (\bar{\mathfrak D}^2 -8R) \bar W^2 \qquad \left(\Rightarrow \  \bar U = ({\mathfrak D}^2 -8\bar R)W^2\right)\, .
\ee
The final result is an action of the form
\be\label{generalsugra}
I=  \int\! d^8 z\, E\, \frac{16W^2 \bar W^2}{U\bar U} \,
L(U,\bar U, {\mathfrak D}W)\, ,
\ee
where ${\mathfrak D}W = {\mathfrak D}^\alpha W_\alpha$ is real as a consequence of the super-Bianchi identities; it is the generalization of the flat superspace scalar superfield $\mathbb D$.

For superModMax coupled to supergravity we have
\be
L = -\frac 1{16}(\cosh\gamma) (U+ \bar U) + \frac 1{8}(\sinh\gamma)\sqrt{U\bar U}\, ,
\ee
and  the action \eqref{generalsugra} for this choice of $L$ can be rewritten as a curved superspace generalization of (\ref{SMM}):
\be\label{sugra+SMMBI}
I= (\cosh\gamma)\left(\frac14 \int \! \d^6\zeta_L\,{\mathcal E}\, { W}^2 + c.c. \right) + 2(\sinh\gamma) \int d^8z\, E\, \frac{W^2\bar{ W}^2}{\sqrt{U\bar U}}\, .
\ee
The first term is an integral over chiral superspace (plus complex conjugate) with the standard chiral superspace measure ${\mathcal E}$ (see \cite{Wess:1992cp}); this is the action of superMaxwell coupled to supergravity and, as is well-known, it is super-Weyl invariant. As we shall see, this is also a property of superModMax coupled to supergravity,
which means that it actually couples only to the fields of {\it conformal} supergravity.

\subsubsection{Super-Weyl and superconformal invariance}

The super-Weyl transformations of the chiral and full superspace measures are
\be
E \to e^{(\Upsilon+\bar\Upsilon)} E\, ,  \qquad \mathcal{E} \to e^{3\Upsilon} \mathcal{E}\, ,
\ee
where the scalar superfield parameter $\Upsilon$ is chiral ($\mathfrak{D}_{\dot\alpha}\Upsilon=0$).  In addition
\be
W_\alpha \to  e^{-\frac 32{\Upsilon}}W_\alpha\, \qquad \left(\Rightarrow \ W^2 \to  e^{-\frac 32{\Upsilon}}W^2\right).
\ee
which confirms the super-Weyl invariance of  $\mathcal{E} W^2$ (and hence of superMaxwell coupled to supergravity).
We also have
\be
\label{supW=bD2-R=S}
(\bar{\mathfrak{D}}\bar{\mathfrak{D}}- 8{R})  \to   e^{-2{\Upsilon}}(\bar{\mathfrak{D}}\bar{\mathfrak{D}}- 8{R}) e^{\bar{\Upsilon}}\, .
\ee
It follows that
\be
 U \; \to  \; e^{-2({\Upsilon}+\bar{\Upsilon})}   U  +  \dots \; .
\ee
where the terms omitted all involve a factor of $W_\alpha$, so that
\be
\frac {W^2}{\sqrt{U}} \,\,\to \,\,e^{\bar{\Upsilon}-2\Upsilon}\,\frac {W^2}{\sqrt{U}}  \qquad \left(\Rightarrow \ \frac {W^2\bar W^2}{\sqrt{U\bar U}}\,\, \to  \,\, e^{-(\Upsilon+\bar\Upsilon)}\,\frac {W^2\bar W^2}{\sqrt{U\bar U}} \right)\, .
\ee
We now see that the interaction term in \eqref{sugra+SMMBI} is also super-Weyl invariant, so
superModMax coupled to supergravity is super-Weyl invariant. This implies that its restriction to flat superspace is superconformal invariant because this is the residual symmetry of the combined super-Weyl and superdiffeomorphism invariance of the supergravity coupled action when the supergravity fields are restricted to their vacuum values.

\subsection{SuperModMax in components}\label{sec:cpts}

To pass to the spacetime component form of the superModMax Lagrangian density \eqref{SMM}, it is convenient to first rewrite
it as
\be\label{LSMM=}
{\mathcal L}_{SMM}= (\cosh\gamma) \mathcal{L}_{SM}
+ \frac {{\rm sinh}\gamma} 2 \int\! d^2\theta\,d^2\bar\theta \, \frac { W^2}
{\sqrt{ \left(-\frac14 \bar{{\mathcal D}}^2\bar{W}^2 \right)}} \, \frac {\bar{W}^2}
{\sqrt{\left(-\frac14 {\mathcal D}^2 W^2 \right)}}\,.
\ee
This is motivated by the fact that both $W^2$ and $\sqrt{-\tfrac14\bar{\mathcal D}^2\bar{W}^2}$ are
chiral scalar superfields
and hence so is their quotient, which has the following component expansion:
\be
\frac {W^2} {\sqrt{- \frac14\bar{\mathcal D}^2\bar{W}^2}}= e^{i\theta\sigma^m\bar{\theta}\partial_m}\left({\mathfrak s} +
\sqrt{2}\,{\theta}^\alpha{\cal V}_\alpha\, +\, {\theta}{\theta}\, {\mathfrak F}\right)\; ,
\ee
with
\bea\label{SV}
{\mathfrak s} &=& - \frac {\lambda^2} {\sqrt{{\bar F}}} \, , \\
{\cal V}_\alpha &=& \frac {1} {\sqrt{{\bar F}}} \left(\chi_\alpha +  \frac {i\lambda^2}{2{\bar F}}({\sigma}^m\partial_m\bar{\chi})_\alpha\right) \, , \label{CV} \\
{\mathfrak F} &=&  \frac {1} {\sqrt{{\bar F}}}\left( {F}+
 \frac {i(\chi{\sigma}^m\partial_m\bar{\chi}) }{2{\bar F}}
 - \lambda^2
 \frac {\Box(\bar\lambda^2)}{2{\bar F}} +\frac 3 8  \, \lambda^2\,  \frac { \partial_m\bar{\chi}\tilde{\sigma}^m {\sigma}^n\partial_n\bar{\chi}}{{\bar F}^2}  \right)\label{frakF}\, ,
\eea
where $\lambda^2$ as defined in \eqref{lam2}, and $\chi_\alpha$ and $F$ are given by \eqref{chi} and \eqref{F=2} respectively. In terms of these components, the superModMax action takes the following form
\bea\label{SMM=comp}
I_{SMM} &=& {\rm cosh}\gamma\, I_{SM} +\frac12 \sinh\gamma\,  I_{\rm int}  \\
&=& \int\! d^4x \left\{\frac12(\cosh\gamma) \Re(F) + \frac12(\sinh\gamma)\left(-\partial_m{\mathfrak s}\partial^m\bar{{\mathfrak s}} - \ii {\cal V}\sigma^m\partial_m\bar{{\cal V}}+
{\mathfrak F}\bar{{\mathfrak F}}\right)\right\}\, . \nonumber
\eea
The equation of motion of the auxiliary field $D$ is
\be\label{D=Eq}
D= - \sqrt{2} \,\sinh\gamma \left[\cosh\gamma +2(\sinh\gamma)\Re\left(\frac {\delta I_{\rm int}}{\delta F}\right)\right]^{-1}
\Im  \left( \lambda^\alpha    \frac {\delta I_{\rm int}\;} {\delta \chi^\alpha}\right)\,.
\ee
Although the right-hand side depends on $D$, it is proportional to a nilpotent fermionic field bi-linear, which allows
an order by order solution that must terminate. The solution to second order in $\lambda$ (and $\bar\lambda$) is
\be\label{D=2f}
D= - \frac {\sinh \gamma} {\left[\cosh \gamma + (\sinh \gamma) \cos\varphi\right]\sqrt{S^2+P^2}}\; ( \lambda \sigma^m \bar{\lambda})\,\tilde F_m{}^n\partial_n\varphi \ + \dots
\ee
where $\varphi$ is the phase of $S+iP$ and the dots stand for terms that are higher order in $(\lambda,\bar\lambda)$.
This solution implies that
\be
\chi^\alpha =- \frac 1{\sqrt{ 2}} \, F_{mn}(\lambda\sigma^{mn})^{\alpha} + \dots
\ee
These results are sufficient to determine the component form of the superModMax Lagrangian density to
second order in fermions. We find that
\bea\label{LSMM=comp1}
{\cal L}_{SMM} &=& \cosh \gamma \left[S  + \Im (\lambda \sigma^m\partial_m\bar{\lambda})\right]
+ \sinh\gamma \left[\sqrt{S^2+P^2} + (\cos\varphi)\Im\left(\lambda \sigma^m\partial_m\bar{\lambda}\right)\right] \nonumber \\
&& \ + \ \frac{\sinh\gamma}{2\sqrt{S^2+P^2}} \,(\lambda\sigma^{n}\bar{\lambda})\,
\tilde F_{pn} \tilde F^{pm}\, \partial_m\varphi\  + \dots
\eea
This preserves parity, as expected; in particular the term in second line is parity-even because both $\varphi$ and
the fermion bilinear $\lambda\sigma^m\bar\lambda$ are parity-odd.

Another feature of  the last term of \eqref{LSMM=comp1} is that the derivative of $\varphi$ implies the presence
of a time derivative of the electric field. Normally, this would indicate that we have a higher-derivative theory with a different canonical structure to super-Maxwell electrodynamics and consequent violations of semi-classical unitarity. However, this (and all other) higher-derivatives appear in nilpotent terms in the action,
and in such cases it is believed that they will not affect the canonical structure because the gauge field equations
can be solved by an order-by-order procedure that must terminate, as argued above for the auxiliary field equation
(see \cite{Kuzenko:2005wh,Farakos:2013fne,Fujimori:2017kyi} for a related complementary discussion of this issue).
As we shall see in Section \ref{sec:VA}, this intuition can be confirmed by means of a nonlinear field redefinition.

Finally, it must be noted that the last term of \eqref{LSMM=comp1} is singular at $S^2+P^2=0$; this is an expected feature of the non-analyticity of the ModMax Lagrangian density at $S^2+P^2=0$, which implies that the ModMax EL equations are not defined for Lagrangian field configurations with $S^2+P^2=0$. However, the Hamiltonian field equations of ModMax are well-defined for those Hamiltonian configurations that imply  $S^2+P^2=0$ \cite{Bandos:2020jsw}. There is therefore reason to expect the same to be true of superModMax, but we leave this to a future investigation.


\section{Born-Infeld-like extension of superModMax}

To apply the same supersymmetrisation procedure to the Born-Infeld-like generalization of ModMax with Lagrangian density of \eqref{BIgammaL*},
 we must start from a superfield Lagrangian of the form (\ref{Lchoice}) with
\be\label{SBI1}
L({\mathbb S},{\mathbb P})= T - \sqrt{T^2 -2T\left[(\cosh\gamma){\mathbb S} +  (\sinh\gamma)\sqrt{{\mathbb S}^2 +{\mathbb P}^2}\right] - {\mathbb P}^2}\, .
\ee
By `adding and subtracting' a free field $(\cosh\gamma)\mathbb S$ term, the resulting superfield Lagrangian density 
can be written in the form 
\be\label{SBI}
\mathcal L_{(\gamma SBI)}= \frac{\cosh\gamma}{4}\left(\int\! \d^2\theta\, { W}^2+{\rm c.c.}\right) +
\frac 14 \int\! d^2\theta d^2\bar\theta\,  {W^2{\bar{ W}}^2 }\, K({\mathbb S},{\mathbb P})\, , 
\ee
where the first, chiral superspace, term is found from the `added' free-field term as spelled out for super-Maxwell in \eqref{Maxchoice}, and 
the `subtracted' free-field term combines with the expression of \eqref{SBI1} to yield the second, full superpace, term with 
\be\label{Lnonli}
K({\mathbb S},{\mathbb P}) = \frac{T - \sqrt{T^2 -2T\left(\cosh\gamma\,{\mathbb S} +  \sinh\gamma\,\sqrt{{\mathbb S}^2 +{\mathbb P}^2}\right) -{\mathbb P}^2}-\cosh\gamma\, {\mathbb S}}{({\mathbb S}^2 + \mathbb{P}^2)}\,.
\ee
In the $T\to\infty$ limit we recover the superModMax superfield Lagrangian density of \eqref{SMM}.  
For $\gamma=0$ we recover the BI superfield Lagrangian density of \cite{Deser:1980ck,Cecotti:1986gb,Bagger:1996wp} (see also e.g. \cite{Rocek:1997hi,Bellucci:2000bd,Ivanov:2001gd,Bandos:2001ku,Bellucci:2015qpa,Cribiori:2018dlc} for further study of supersymmetry properties of this theory and its alternative formulations). For example, we may rewrite \eqref{Lnonli} for $\gamma=0$ as
\be\label{KBG}
K|_{\gamma=0}=\frac{1}{T-\mathbb S+\sqrt{T^2-2T\,\mathbb S - \mathbb P^2}}  \, ,
\ee
which is the nonlinear part of the Bagger-Galperin super-BI superspace Lagrangian \cite{Bagger:1996wp}.

\subsection{Duality invariance}

To verify duality invariance of the BI-like generalization of ModMax, it is convenient to use,  following \cite{Kuzenko:2000tg,Kuzenko:2000uh}, the flat superspace counterparts of the variables \eqref{defU}:
\be\label{uu}
u:= {\mathcal D}^2 W^2=-8(\mathbb S-\ii\mathbb P)\, , \qquad \bar u := \bar {\mathcal D}^2 \bar W^2=-8(\mathbb S+\ii\mathbb P)\,\, .
\ee
In this notation, the function $L(\mathbb S,\mathbb P)$ of \eqref{SBI1} takes the form\footnote{To prove the duality invariance there is no need to split the Lagrangian density into free-field and interaction terms, as was done in \cite{Kuzenko:2000tg,Kuzenko:2000uh}.}
\be\label{Lnew}
L(u,\bar u) = T - \sqrt{T^2 +\frac T8\left[(\cosh\gamma)(u+\bar u) -  2(\sinh\gamma)\sqrt{u\bar u}\right] +\frac{(u-\bar u)^2}{256}}\, ,
\ee
and the chiral superfield $M_\alpha$ of \eqref{SDC} is
\be\label{MBIMM1}
M_\alpha= 16{\ii}\, W_\alpha \bar {\mathcal D}^2\left\{\bar W^2\left[\frac L{u\bar u}+{\mathcal D}^2\left(\frac{W^2}{\bar u}\partial_u\left(\frac{L}{u}\right)\right)\right]\right\}\, .
\ee
Using these results, the duality invariance condition \eqref{SDC} for the generic $L(u,\bar u)$ becomes
\bea\label{SDCBIMM1}
0 &=& \Im\int\! d^4x\left\{ \int\!d^2\theta \,W^2
+(32)^2\int\! d^2\theta\,d^2\bar\theta \,W^2\bar W^2\,\frac{L_u^2} {\bar u}\right\} \nonumber\\
&=&\Im\int\! d^4x\,\left\{-\frac u 4
+(32)^2\,\int d^2\theta\,d^2\bar\theta \,W^2\bar W^2\,\frac {L_u^2}{\bar u}\right\}\, .
\eea
This condition is satisfied because for the BI-like Lagrangian density \eqref{Lnew} the following identity holds
\be\label{Eq21}
\Im\,u\left[\left(16\,L_u\right)^2-1\right]=0\,.
\ee
This equation is formally identical to the condition on the {\it bosonic} Lagrangian density required for duality invariance of its EL equations \cite{Gibbons:1995cv,Gaillard:1997rt,Gaillard:1997zr,Kuzenko:2000tg}.

\section{Taming the higher derivatives}\label{sec:VA}

We will now show, following a  procedure described in \cite{Cribiori:2018dlc} for a superfield formulation of super-BI theory\footnote{We are grateful to Fotis Farakos for a detailed explanation of this procedure.},
how one can rewrite the superfield Lagrangian \eqref{SMM} in a Volkov-Akulov-like form by effectively
converting the superfield strength $W_\alpha$ into constrained superfields each of which contains a single independent component of the vector multiplet. Upon this (step-by-step) field redefinition, the $\mathcal N=1$ supersymmetry becomes non-linearly realized on the new component fields.\footnote{The procedure described below has its roots in a general relation between linear and non-linear realizations of supersymmetry put forward in \cite{Ivanov:1978mx,Ivanov:1982bpa}.}  This is the price one pays for removing the higher derivative terms from the action of any generic non-linear supersymmetric electrodynamics theory.

We take as our starting point the Lagrangian density \eqref{Lchoice}
 with $L$ a function of the superfields
$(\mathbb S,\,\mathbb P)$ and $\mathbb D$ that were defined in \eqref{SPSF}:
\be\label{SMME}
\mathcal L =\int\! d^2\theta\, d^2\bar\theta \frac{16 W^2{\bar{ W}}^2 }{{\mathcal D}^2 W^2\,\bar {\mathcal D}^2{\bar{ W}}^2}\,L(\mathbb S,\mathbb P,\mathbb D)\,.
\ee

\subsection{Step 1: nonlinear superfield redefinition}
The first step of the procedure is to introduce the spinor superfield
\be\label{Gamma}
\Gamma_\alpha :=-2\frac{{\mathcal D}_\alpha W^2}{{\mathcal D}^2 W^2}\,,
\ee
which is defined only if ${\mathcal D}^2W^2 \ne 0$; we postpone discussion of the implications of this restriction.
Because $(W^2)W_\alpha\equiv 0$, the spinor superfield $\Gamma_\alpha$
satisfies
\be\label{Grelate}
\Gamma^2 = -4 \frac{W^2}{{\mathcal D}^2 W^2}   \quad \Rightarrow\quad  {\mathcal D}_\alpha \Gamma^2=2\Gamma_\alpha \quad \Rightarrow\quad  {\mathcal D}^2\Gamma^2=-4\, .
\ee
From \eqref{Gamma} one can also derive the following additional relations:
\be\label{SW}
{\mathcal D}_\alpha \Gamma_\beta=-\varepsilon_{\alpha\beta}\,,\qquad
\bar {\mathcal D}_{\dot \alpha}\Gamma_{\beta}=-2i \Gamma^{\rho}\,\sigma^m_{\rho\dot\alpha}\,\partial_m\Gamma_{\beta}\,.
\ee
These relations tell us that $\Gamma_\alpha$ is the (goldstino) constrained superfield first introduced in
\cite{Ivanov:1978mx} and further elaborated in \cite{Samuel:1982uh} . It has the single independent component
\be\label{gammaf}
\zeta_\alpha=\Gamma_\alpha|=\frac{\chi_\alpha}{\sqrt {2}\,F}\,,
\ee
where $\chi_\alpha$ and $F$ were defined, respectively, in \eqref{chi} and \eqref{F=2}. The spinor $\zeta_\alpha$ transforms non-linearly under supersymmetry
\be\label{susygamma}
\delta\,\zeta_\alpha=\epsilon_\alpha-2\ii\,\zeta\,\sigma^m\bar \epsilon\,\partial_m\zeta_\alpha\,.
\,\ee
The relation of $\zeta$ to the original Volkov-Akulov goldstino \cite{Volkov:1972jx,Volkov:1973ix}, which we shall call $\nu_\alpha$,
is by a particular invertible nonlinear field redefinition ($\zeta_\alpha=\nu_\alpha + O(\nu^3)$) \cite{Ivanov:1978mx,Samuel:1982uh} which we now aim to explore at the level of superfields\footnote{For a review of relations between different constrained superfields, which describe the Volkov-Akulov goldstino,  and the original VA construction see e.g. \cite{Bandos:2016xyu}.} by the introduction of a ``Volkov-Akulov'' superfield $\Lambda_\alpha$ for which $\Lambda_\alpha \big| = \nu_\alpha$. This superfield satisfies the constraints
\cite{Ivanov:1978mx}  (see also \cite{Wess:1992cp,Ivanov:2016lha,Cribiori:2018dlc})
\be\label{Lambda}
{\mathcal D}_\alpha \Lambda_\beta=-\varepsilon_{\alpha\beta}+i\sigma^m_{\alpha\dot\beta}\bar\Lambda^{\dot\b}\,\partial_m\Lambda_\beta\,,\qquad
\bar{\mathcal D}_{\dot \alpha}\Lambda_{\beta}=-i \Lambda^{\rho}\,\sigma^m_{\rho\dot\alpha}\,\partial_m\Lambda_{\beta}\,.
\ee
These constraints imply that the leading component $\nu_\alpha$ of $\Lambda_\alpha$ is its only independent one; all others
are determined by $\nu_\alpha$ and its derivatives.

The superfields $\Gamma_\alpha$ and $\Lambda_\alpha$ are  related by a superfield version of the invertible nonlinear redefinition that relates their leading components; the map from $\Lambda$ to $\Gamma$ is \cite{Cribiori:2018dlc}
\be\label{GL}
\Gamma_\alpha=-2\frac{{\mathcal D}_\alpha \bar {\mathcal D}^2(\Lambda^2\bar\Lambda^2)}{{\mathcal D}^2\bar {\mathcal D}^2(\Lambda^2\bar\Lambda^2)}\,,
\ee
which implies the identities
\be
\Gamma^2\bar\Gamma^2 \equiv \Lambda^2\bar\Lambda^2 \equiv
16\frac{ W^2{\bar{ W}}^2 }{{\mathcal D}^2 W^2\,\bar {\mathcal D}^2{\bar{ W}}^2}\,.
\ee
Other useful identities are
\bea\label{identities}
\bar{\mathcal D}^2(\Lambda^2\bar\Lambda^2)\propto \Lambda^2 \qquad &\Rightarrow& \   \left(\bar {\mathcal D}^2(\Lambda^2\bar\Lambda^2)\right)^2=0\, , \nonumber \\
-\frac 14 \Gamma^2\equiv\frac{W^2}{{\mathcal D}^2W^2}\equiv \frac{\bar {\mathcal D}^2(\Lambda^2\bar\Lambda^2)}{{\mathcal D}^2\bar {\mathcal D}^2(\Lambda^2\bar \Lambda^2)}\qquad &\Rightarrow& \ \bar {\mathcal D}^2(\Lambda^2\bar\Lambda^2) W_\alpha=0\,.
\eea
These relations imply that
\be\label{VA}
\int\! d^4 x\,  d^2\theta d^2\bar\theta \frac{16\, W^2{\bar{ W}}^2 }{{\mathcal D}^2 W^2\,\bar{\mathcal D}^2{\bar{ W}}^2}=
\int\! d^4x \,d^2\theta d^2\bar\theta \,\Lambda^2\bar\Lambda^2 =\int\! d^4 x \, \det{\mathbb E}_{m}^{\ a} \big|\,,
\ee
where
\be
{\mathbb E}_m^{\ a} =\delta_m^a+\ii \Lambda\, \sigma^a\,\overleftrightarrow{\partial_m}\, \bar\Lambda\, ,
\ee
which tells us that $\det E_m^{\ a} \equiv \det{\mathbb E}_{m}^{\ a} \big|$ is  (minus) the Volkov-Akulov Lagrangian density for
the goldstino field $\nu$. For later use, note the superfield identities
\be\label{sfidentity}
\Lambda^2\bar\Lambda^2\,\det {\mathbb E}_m^{\ a}\equiv \Lambda^2\bar\Lambda^2\,, \qquad
\Lambda^2\bar\Lambda^2\,\mathbb E_a^{-1\,m}\,\equiv \Lambda^2\bar\Lambda^2\,\delta^m_a\,.
\ee
The action for the Lagrangian density \eqref{SMME} can now be put into the form
\be\label{SMME1}
I=\int\! d^4 x \, \mathcal L=\int\! d^4 x \,d^2\theta d^2\bar\theta\,\Lambda^2\, \bar\Lambda^2\, L(\mathbb S,\mathbb P,\mathbb D)\, .
\ee

\subsection{Step 2: integration of the Grassmann variables}

To proceed, it is convenient  to define, with the use of \eqref{D^2W^2}, the new superfields
\bea\label{hatSP}
\widehat{\mathbb S}&=& {\mathbb S}+\frac 18(W^\alpha {\mathcal D}^2 W_\alpha+c.c.)=-\frac 1{4}\,{\mathbb F}_{mn}{\mathbb F}^{mn}+\frac 12 \mathbb D^2\,,\nonumber\\
 \widehat{\mathbb P}&=&{\mathbb P}-\frac {\ii}8(W^\alpha {\mathcal D}^2 W_\alpha-c.c.)=-\frac 1{4}\,{\mathbb F}_{mn}\widetilde{\mathbb F}^{mn}\,,
\eea
where the superfields
\be\label{mF}
{\mathbb F}_{mn}=2\partial_{[m}\,{\mathbb A}_{n]}, \qquad {\mathbb A}_m=
{\frac {1} 4 }
({\mathcal D}^\alpha\sigma_{m\a\dot\a}\bar {\mathcal D}^{\dot\a}-\bar {\mathcal D}_{\dot \a}\tilde \sigma_m^{\dot\a\a}{\mathcal D}_\a)\,V\,,
\ee
have leading components $F_{mn}$ and $A_m$ respectively; the superfield $V$ is the prepotential introduced in \eqref{WV}.
This allows us to write
\be\label{L2}
 L(\mathbb S,\mathbb P,\mathbb D)=L\left(\widehat{\mathbb S},\widehat{\mathbb P},{\mathbb D}\right)+O(W_\alpha,\bar W_{\dot\alpha}),
 \ee
where $O(W_\alpha,\bar W_{\dot\alpha})$ stands for terms that are annihilated by $\Lambda^2\bar\Lambda^2\sim W^2\bar W^2$.

In \cite{Cribiori:2018dlc} it was shown that, for any real superfield $L$,
\be\label{integ}
\int\! d^4 x \,d^2\theta d^2\bar\theta\,  \Lambda^2\bar\Lambda^2\, L=\frac 1 {16}\int\! d^4 x\,\left(\det E_m^{\ a}\right) \left[\Pi^2\bar\Pi^2(\Lambda^2\bar\Lambda^2\,L)\right]_{\theta,\bar\theta=0}\,,
\ee
where
\be\label{Pi}
\Pi_\alpha={\mathcal D}_\alpha-i\sigma^m_{\alpha\dot\alpha}\bar\Lambda^{\dot\alpha}\,\partial_m\, ,  \qquad \bar\Pi_{\dot\alpha}=(\Pi_\alpha)^*=\bar {\mathcal D}_{\dot\alpha}+i\Lambda^{\alpha}\sigma^m_{\alpha\dot\alpha}\,\partial_m\, ,
\ee
are covariant derivatives forming  the (anti)commutative algebra\footnote{The second bracket is zero because $\Pi_\alpha\, \mathbb E^{-1\,m}_a=-\ii \sigma^m_{\alpha\dot\alpha}\,\Pi_a\,\bar\Lambda^{\dot\a}$.}
\be\label{Pia}
\{\Pi_\alpha,\Pi_\beta\}=0,\qquad \{\Pi_{\alpha},\bar\Pi_{\dot\beta}\}=0,\qquad [\Pi_\alpha,\Pi_a]=0\,, \qquad \Pi_a\equiv {\mathbb E}^{-1 \,m}_a\partial_m
\ee
They act on $\Lambda_\alpha$ as follows
\be\label{PiLambda}
\Pi_{\alpha}\,\Lambda_{\beta}=-\varepsilon_{\a\b}\,,\qquad \bar\Pi_{\dot\alpha}\Lambda_\beta=0\,
\quad \left(\Rightarrow \ \Pi^2\Lambda^2=-4\right)\,.
\ee
These are just the goldstino superfield constraints of \eqref{Lambda}.
An instructive identity is
\be\label{Pi E-1}
\Pi_{[a}\, \mathbb E_{b]}^{-1\,m}\equiv -2\ii\,\left(\Pi_{[a}\Lambda\,\sigma^c\Pi_{b]}\bar\Lambda\right)\,\mathbb E_{c}^{-1\,m} 
\ee
because this implies
\be
[\Pi_a,\Pi_b]=-4\ii\,\left(\Pi_{[a}\Lambda\,\sigma^c\Pi_{b]}\bar\Lambda\right)\,\Pi_c\, , 
\ee
which tells us that the algebra of the covariant derivatives $\Pi_A$ determines an unconventional flat superspace with a non-vanishing torsion tensor $T_{AB}{}^C$ for which the only non-zero component is  $T_{ab}^c=-4\ii\,\left(\Pi_{[a}\Lambda\,\sigma^c\Pi_{b]}\bar\Lambda\right)$.

Using the  relations \eqref{Pia},
one proves that the operator $\frac 1{ 16} \Pi^2\bar\Pi^2\,\Lambda^2\bar\Lambda^2$ is a projector in the sense that
\be\label{2timesPi}
\frac 1{16^2} \Pi^2\bar\Pi^2[\Lambda^2\bar\Lambda^2\Pi^2\bar\Pi^2(\Lambda^2\bar\Lambda^2\,L)]=\frac 1{16} \Pi^2\bar\Pi^2(\Lambda^2\bar\Lambda^2\,L)\,.
\ee
This means that if $L=f(M)$, some function of another superfield $M$, then (as proved in eqs. \eqref{fM+X}-\eqref{M+} of the Appendix)
\be\label{L=fM}
\frac 1{16} \Pi^2\bar\Pi^2\Big(\Lambda^2\bar\Lambda^2\,f(M)\Big)=f\left(\frac 1{16} \Pi^2\bar\Pi^2(\Lambda^2\bar\Lambda^2\,M)\right)\,.
\ee
We thus arrive at a Lagrangian density of the following form
\be\label{LDVA}
\mathcal L= \left(\det E_m^{\ a}\right)\,L(\mathcal S, \mathcal P, \mathfrak D)\vert_{\theta,\bar\theta=0}\,,
\ee
where
\be\label{mathcal SP}
\mathcal S=-\frac 14 \mathcal F_{ab}\mathcal F^{ab}+\frac {{\mathfrak D}^2}2, \qquad \mathcal P=-\frac 14 \mathcal F_{ab}\tilde{\mathcal F}^{ab}\,,
\ee
with
\be\label{calD}
\mathfrak D =\frac 1{16} \Pi^2\bar\Pi^2(\Lambda^2\bar\Lambda^2\,\mathbb D)\, , \quad
\ee
\be\label{calF}
\mathcal F_{ab} =\frac 1{16} \Pi^2\bar\Pi^2(\Lambda^2\bar\Lambda^2\,\mathbb F_{ab})\, , \quad {\mathbb F}_{ab}=2{\mathbb E}_a^{-1\,m}{\mathbb E}_b^{-1\,m}\partial_{[m} \mathbb A_{n]} \,.
\ee

\subsection{Step 3: partial fixing of supergauge symmetry}

The next step is to use \eqref{Pia}, \eqref{Pi E-1} and the second identity in \eqref{sfidentity} to
rewrite $\mathcal F_{ab}$ of \eqref{calF} as follows:
\bea\label{calF1}
\mathcal F_{ab}&=&\frac 1{8} \Pi^2\bar\Pi^2(\Lambda^2\bar\Lambda^2\,\partial_{[a}\mathbb A_{b]})=
\frac 1{8} \Pi^2\bar\Pi^2\left(\Lambda^2\bar\Lambda^2\,\mathbb E^{-1\,m}_{[a}\mathbb E^{-1\,n}_{b]}\partial_{m}\mathbb A_{n}\right)\nonumber\\
&=&\frac 1{8} \Pi^2\bar\Pi^2\Pi_{[a}\left(\Lambda^2\bar\Lambda^2\,\mathbb E^{-1\,n}_{b]}\mathbb A_{n}\right) - \frac 18\Pi^2\bar\Pi^2\left((\Pi_{[a}\mathbb E^{-1\,n}_{b]})\mathbb A_{n}\Lambda^2\bar\Lambda^2)\right)\nonumber\\
&+&\frac 1{8} \Pi^2\bar\Pi^2\left(\mathbb E^{-1\,n}_{[a}\mathbb A_{n}\Pi_{b]}(\Lambda^2\bar\Lambda^2)\right)\\
&=&\frac 1{8} \mathbb E^{-1\,m}_{[a}\mathbb E^{-1\,n}_{b]}\partial_m\left(E^c_n\Pi^2\bar\Pi^2\left(\Lambda^2\bar\Lambda^2\,\mathbb A_{n}\right)\right)+\frac 1{8} \Pi^2\bar\Pi^2\left(E^{-1\,n}_{[a}\mathbb A_{n}\Pi_{b]}(\Lambda^2\bar\Lambda^2)\right)\, .  \nonumber
\eea
This is equivalent to
\be\label{calFfinal}
\mathcal F_{ab}=2\, \mathbb E^{-1\,m}_{[a}\mathbb E^{-1\,n}_{b]}\partial_m\mathcal A_n+\frac 1{8} \Pi^2\bar\Pi^2\left(E^{-1\,n}_{[a}\mathbb A_{n}\Pi_{b]}(\Lambda^2\bar\Lambda^2)\right),
\ee
where
\be\label{calA=}
\mathcal A_n= \frac 1{16} \mathbb E_n^{\ a}\,  \Pi^2\bar\Pi^2(\Lambda^2\bar\Lambda^2\,\mathbb  E_{a}^{-1\,m} \mathbb A_m)\,=\frac 1{16} \mathbb E_n^{\ a} \,  \Pi^2\bar\Pi^2(\Lambda^2\bar\Lambda^2\,\mathbb A_a).
\ee

All higher derivative terms in the Lagrangian come from the second term in \eqref{calFfinal}. We will now show that this term can be gauged away by a partial gauge fixing of the supergauge invariance of the theory. To this end, we
observe that the expression for $\mathcal F_{ab}$ in \eqref{calFfinal} is manifestly invariant under the following supergauge transformations with chiral superfield parameter $\Phi$:
\be\label{V'}
\delta_\Phi V = \ii(\Phi-\bar \Phi) \qquad \Rightarrow \quad \delta_\Phi \mathbb A_a = \partial_a(\Phi+\bar\Phi)\,,
\ee
but the vector field $\mathcal A_m$ does not transform in the conventional way; instead
\bea\label{Vtransform}
\delta \mathcal A_n&=&\frac 1{16}\partial_n \left(\Pi^2\bar\Pi^2\left[\Lambda^2\bar\Lambda^2\,(\Phi+\bar\Phi)\right]\right)-\frac 1{16}\mathbb E_n^{\ a}\Pi^2\bar\Pi^2[\Pi_a(\Lambda^2\bar\Lambda^2)\,(\Phi+\bar\Phi)].
\eea
To resolve this difficulty, the authors of \cite{Cribiori:2018dlc} proposed to impose on $V$ (and hence on $\mathbb A_m$) a Wess-Zumino-like gauge introduced in \cite{Komargodski:2009rz}, where it was shown that there exists a $\Phi$ such that
\be\label{XV=0}
XV=0,
\ee
where $X$ is a nilpotent chiral superfield,  which in our case is
\be\label{X}
 X=-\frac 14\bar D^2(\Lambda^2\bar\Lambda^2), \quad \Rightarrow \ X^2=0\,, \quad  X\bar X= \Lambda^2\bar\Lambda^2\,.
\ee
A solution $V= V^\prime$ of \eqref{XV=0} remains a solution under the transformation
\be\label{resi}
V^\prime \to V^\prime +\ii(\Phi^\prime-\bar{\Phi}^\prime)
\ee
provided the chiral scalar superfield parameter $\Phi^\prime$ satisfies the constraint
\be\label{checkPhi}
X\Phi^\prime=X\overline{\Phi}^\prime\,.
\ee
In this (partial) gauge, the gauge potential $\mathcal A_n$ of \eqref{calA=} reduces to
\be\label{calA=1}
{{\mathcal A}}^\prime_n= \frac 1{16} \mathbb E_n^{\ a}  \Pi^2\bar\Pi^2(\Lambda^2\bar\Lambda^2\,  {\mathbb A}^\prime_a)\, ,
\ee
which was shown in \cite{Cribiori:2018dlc} to transform as in  \eqref{Vtransform}  but without the second term; i.e. it transforms as
a genuine gauge potential under the residual gauge transformations \eqref{resi} subject to \eqref{checkPhi}:
\be\label{Vtransformcheck}
\delta {\mathcal A}^\prime_n = \partial_n\, \phi\, , \qquad
\phi= \frac{1}{16} \Pi^2\bar\Pi^2\left[\Lambda^2\bar\Lambda^2\left(\Phi^\prime+\overline{\Phi}^\prime\right)\right]\, .
\ee
It was also shown in \cite{Cribiori:2018dlc}  that
\be\label{Aconst}
{\mathcal A}^\prime_n={\mathbb A}^\prime_n+ {\cal O}(\Lambda^2)\, ,
\ee
where ${\cal O}(\Lambda^2)$ stand for terms which are at least quadratic in $\Lambda$ and/or $\bar\Lambda$.
It then follows that
\be
\mathbb  E_{a}^{-1\,n} {\mathbb A}^\prime_{n}=\Pi^2\bar\Pi^2(\Lambda^2\bar\Lambda^2 {\mathbb A}^\prime_a)+{\cal O}(\Lambda^2)
\ee
and that the last term in \eqref{calFfinal} vanishes, because ${\cal O}(\Lambda^2)\Pi_{b}(\Lambda^2\bar\Lambda^2)\equiv 0$. What remains of this last term is annihilated by $\Pi^2\bar\Pi^2$,
so we finally arrive at the following relation
\be\label{calFfinal1}
\mathcal F_{ab}=\frac 1{8} \Pi^2\bar\Pi^2(\Lambda^2\bar\Lambda^2\,\partial_{[a}\mathbb A_{b]})=2 \,\mathbb E^{-1\,m}_{[a}\mathbb E^{-1\,n}_{b]}\partial_m{\mathcal A}^\prime_n.
\ee

Substitute \eqref{calFfinal1} into \eqref{mathcal SP} and  \eqref{LDVA},  we  get
\be\label{VALgen}
\mathcal L=(\det{E_m^{\ a}})\,L(\mathcal S|, \mathcal P|, \mathfrak D|)\, ,
\ee
in which
\be\label{EVA}
E_m^{\ a} (x)=\delta^a_m+\ii \nu\sigma^a\,\partial_m\bar\nu-\ii \partial_m\nu\sigma^a\,\bar\nu(x)\, ,
\ee
which implies
\be\label{detE}
\det E_m^{\ a} =1+\ii(\nu\sigma^m\partial_m\bar\nu-\partial_m\nu\sigma^m\,\bar\nu)+ O(\nu^4)\,,
\ee
and
\be\label{SPD}
\mathcal S|=-\frac 14 f_{ab}f^{ab}+\frac 12 \mathfrak D^2|, \qquad \mathcal P|=-\frac 14 f_{ab}\tilde f^{ab}\, ,
\ee
where
\be\label{fab}
f_{ab}(x)=2E_{a}^{-1\,m}E_{b}^{-1\,m}\partial_{[m}\,a_{n]}(x), \qquad a_n(x) :={ {\mathcal A}}^\prime_n|\,.
\ee
At this point one sees that $\mathfrak D|$ can be regarded as a redefined auxiliary field of the vector multiplet which can be integrated out in the class of models discussed in Section \ref{SNED}.

\subsection{Final action and its field equations}

From \eqref{VALgen} we see that the action \eqref{SMME1} has been reduced to one of Volkov-Akulov form:
\be\label{finalA}
I=\int\,d^4x \sqrt{-\det{g}}\,L({\tt s}, {\tt p})\,,
\ee
where $g$ is the VA metric (determined by the VA vierbein of \eqref{EVA})
with components
\be\label{VAmetric}
g_{mn}=E_m^{\ a} E_n^{\ b}\eta_{ab}\,  \qquad \left(E_m^{\ a} =\delta_m^a+\ii \nu\, \sigma^a\,\overleftrightarrow{\partial_m}\, \bar\nu\; , \right) \, ,
\ee
and
\bea
{\tt s} &=& - \frac14 f_{ab}f^{ab}=- \frac 14 f_{mn}f_{pq}\,g^{mp}g^{nq}\,:=- \frac 14 f_{mn}f^{mn}, \nonumber \\
{\tt p} &=& - \frac 14 f_{ab}\tilde f^{ab}=- \frac 14 f_{mn}\tilde f^{mn}\,.
\eea
The construction ensures invariance under the following non-linearly realized supersymmetry transformations:
\bea\label{NSUSY}
\delta \nu_\alpha &=&\epsilon_\alpha+\ii(\epsilon\sigma^m\bar\nu-\nu\sigma^m\bar\epsilon)\,\partial_m\nu_\alpha\,, \nonumber \\
\delta a_m &=&\ii(\epsilon\sigma^n\bar\nu-\nu\sigma^n\bar\epsilon)\,\partial_n a_m +\ii\partial_m(\epsilon\sigma^n\bar\nu-\nu\sigma^n\bar\epsilon)\,a_n\,.
\eea
where $a_n$ is the gauge potential introduced in \eqref{fab}. These transformations may be rewritten as specific general coordinate transformations accompanied by a constant shift of the fermionic field:
\bea
\delta x^m &=&\xi^m(x):=-\ii(\epsilon\sigma^m\bar\nu(x)-\nu(x)\sigma^m\bar\epsilon),\nonumber\\ \delta\nu_\alpha &=&\epsilon_\alpha-\xi^m(x)\partial_m\nu_\alpha, \qquad \delta a_m=-\xi^n\partial_na_m-(\partial_m\xi^n)a_n\,,\nonumber\\
\delta E_m^{\ a} &=&-\xi^n\partial_n E_m^{\ a}-(\partial_m\xi^n)E_n^{\ a}\, .
\eea
The action \eqref{finalA}, which has {\it no higher derivative terms}, may therefore be interpreted as the bosonic nonlinear
electrodynamics with Lagrangian density $\mathcal L= L(S,P)$ in the background  VA metric constructed from the goldstino field
$\nu_\alpha (x)$.

It is instructive to examine the field equation that follow from the action \eqref{finalA}.
The gauge field equation is
\be\label{Aeom}
\partial_m\left(\sqrt{-g}\frac{\partial \mathcal L}{\partial{F_{mn}}}\right)=0\, .
\ee
The (complex) field equation for the spinor field $\nu$ is
\bea\label{nu}
&(\sigma_a\partial_n\bar\nu)_\alpha E_m^{\ a}\,T^{nm}
=0\,,&
\eea
where
\be\label{T}
T^{mn}=-\frac 2{\sqrt{-g}}\frac{\partial\mathcal L}{\partial g_{mn}} ,
\ee
which is the energy-momentum stress tensor of the electromagnetic field in a background with the metric of \eqref{VAmetric}. It is
covariantly conserved as a consequence of the gauge field equation \eqref{Aeom}:
\be\label{nablaT=0}
\nabla_nT^{nm}=0\,,  \qquad
\ee
where $\nabla$ is the usual covariant derivative constructed from the metric.

\subsubsection{Application to superModMax}

The results just obtained apply to generic NEDs in the class to which our supersymmetrization prescription applies. As we saw earlier, this includes superModMax and the generalized supersymmetric BI theory that reduces to superModMax in a weak-field limit.
For superModMax, the change of variables described in detail above leads to the following Lagrangian density:
\be\label{SMM3}
{\mathcal L}_{SMM}=\frac14 \sqrt{-\det{g}} \left(-{\cosh\gamma}\, f_{ab}f^{ab}+\sinh\gamma\,\sqrt{(f_{ab}f^{ab})^2+(f_{ab}\tilde f^{ab})^2}\right).
\ee
and the energy-momentum tensor \eqref{T} which appears in the fermionic equations of motion is
\be\label{TMM}
T^{mn}= \left(f^{m}{}_{p}f^{np}-\frac 14 g^{mn} f_{pq}f^{pq}\right)\left(\cosh\gamma+\sinh\gamma\frac{S}{\sqrt{S^2+P^2}}\right)\,.
\ee
Notice that this result applies even for $\gamma=0$, i.e. super-Maxwell, in which case we have managed to express a free-field theory with unbroken supersymmetry in a form in which it is apparently interacting and for which supersymmetry is apparently broken!

\subsection{Supersymmetry: unbroken or broken?}

The paradox of nonlinearly realized supersymmetry for a theory with unbroken supersymmetry applies generally (see \cite{Ivanov:1982bpa} for a related discussion). Its resolution
is that the goldstino field equation \eqref{nu} is identically zero
in the bosonic vacuum in which $T^{mn}=0$, i.e. when the electromagnetic field is zero. This is a consequence of the fact that the goldstino kinetic term supplied by the action \eqref{detE} comes with a factor of $L({\tt s},{\tt p})$, which is {\it zero} in the bosonic vacuum\footnote{For the same reason, the goldstino field  $\nu_\alpha$ does not have its canonical dimension.}. Indeed, if this were not the case, then we would be able to conclude that supersymmetry is spontaneously broken. So the singular nature of the goldstino equation in the bosonic vacuum is an inevitable consequence of unbroken supersymmetry.

In contrast, when supersymmetry is broken, e.g. by the FI-type mechanism summarized in section \ref{SNED},
the bosonic Lagrangian density is negative in the vacuum, and the above difficulties of interpretation of
the goldstino field equation do not arise, as could be expected since supersymmetry is spontaneously broken.
In such cases, the action \eqref{finalA} makes this manifest.

We conclude by mentioning that another way to get a goldstino kinetic term that is defined in the bosonic vacuum is to
choose the function $L$ in \eqref{Lchoice} to have the form
\be\label{L-zeta2}
L(\mathbb S,\mathbb P, \mathbb D) = {\mathcal L}^{\rm bos}(\mathbb S,\mathbb P) - \zeta^2\, ,
\ee
which may be compared with \eqref{Simplechoice1}, where the FI-type term introduces the constant $\xi$. The new constant $\zeta$ has the same dimensions as $\xi$, and it also leads to a spontaneous breakdown of supersymmetry, but in a very different way. Its effect is to add to the action \eqref{Lchoice} the term
\bea\label{VA1}
-\zeta^2\int d^4x\,d^2\theta d^2\bar\theta \frac{16\, W^2{\bar{ W}}^2 }{{\mathcal D}^2 W^2\,\bar{\mathcal D}^2{\bar{ W}}^2} &=&-\zeta^2\,\int d^4x d^2\theta d^2\bar\theta\,\Lambda^2\bar\Lambda^2\nonumber \\
&=& -\zeta^2\int\,d^4x\, \det E_m^{\ a}\,
\eea
The effect of this term in the action \eqref{finalA} is therefore to add a constant to $L({\tt s}, {\tt p})$:
\be\label{zeta}
{L({\tt s}, {\tt p})} \to L({\tt s}, {\tt p}) - \zeta^2\, .
\ee
Now, even though $L({\tt s}, {\tt p})$ is zero in the bosonic vacuum (if we assume unbroken supersymmetry for $\zeta=0$) this
is no longer true for $\zeta\ne0$, and the would-be goldstino field $\nu$ becomes a {\it bona-fide} one, with canonical dimension after
a rescaling by a power of the constant $\zeta$.

This mechanism for supersymmetry breaking is distinct from the FI mechanism because any sign choice for the FI-type term leads to a positive vacuum energy whereas the sign of the $\zeta^2$ term in \eqref{VA1} must be chosen to ensure this, as we have done. The opposite sign choice would lead to a negative vacuum energy, which is compatible with (spontaneously broken) supersymmetry only in field theories with `ghosts'; in this case, the goldstino would be the ghost since the sign of its (VA) Lagrangian density in the bosonic vacuum changes if $\zeta^2 \to -\zeta^2$ in \eqref{VA1}.

.


\section{Conclusions}

Supersymmetric extensions of nonlinear electrodynamics theories have been intensively studied over the last several decades. It has been understood how to construct these theories, couple them to supergravity, and determine whether they are electromagnetic-duality invariant. These results have often been found on the assumption of a superMaxwell weak-field limit and in the context of an action that is the sum of the chiral superspace integral of superMaxwell and a full-superspace integral for the interactions. However, this free-field/interaction separation fails to make transparent some relations between a given bosonic theory and its
supersymmetric extension. Ideally, one would like a simple prescription that takes the Lagrangian density ${\mathcal L}^{\rm bos}$
of a given bosonic electrodynamics theory, expressed as a function of Lorentz scalars, and uses that function in a
single-term formula for the Lagrangian density of its supersymmetric extension. We have provided just such a prescription
on the assumption that ${\mathcal L}^{\rm bos}$ is a strictly convex function of the electric field. This condition is required
on physical grounds for the purely bosonic theory, but it is also important for supersymmetry because it ensures a unique solution of the
auxiliary field equation in the Maxwell supermultiplet used to construct theories of nonlinear supersymmetric electrodynamics.

We were initially motivated to seek this simple general prescription in order to simplify the construction of a supersymmetric extension of the ModMax electrodynamics and its Born-Infeld-like generalization that we introduced in earlier work \cite{Bandos:2020jsw,Bandos:2020hgy}. It has the additional addvantage of making it more evident why properties of the bosonic theory, such as electromagnetic duality invariance, are inherited by the supersymmetric extension. In the case of superconformal invariance of superModMax electrodynamics, we found it convenient to first  couple to supergravity, which is accomplished by straightforward generalization of the flat-superspace action, and then establish super-Weyl invariance of the result.

We have presented the component action of the superModMax theory up to quadratic order in fermions. Already at this order we confirm the presence of a higher derivative interaction of the electromagnetic fields with the photino field. It has long been understood that this should not be a problem in principle because it should be possible to eliminate the higher-derivative terms from the field equations by an iterative process that must terminate because of the anticommutativity of the photino field. However, this iterative process has never been completely carried out,
and it is restricted to field equations; given that it works, one would expect there to exist a nonlinear field redefinition that removes all higher-derivative terms from the action. We have shown that  this is accomplished by a particular non-linear field redefinition proposed in \cite{Cribiori:2018dlc} and based on ideas in \cite{Ivanov:1978mx,Ivanov:1982bpa}. The final result is surprisingly simple, and it involves a formal
re-interpretation of the photino as  a Volkov-Akulov goldstino. This appears to be the price that must be paid for the explicit elimination of all higher-derivative terms from the action.

A difficulty with the standard Lagrangian formulation of superModMax is that the photino field equations are not defined in the bosonic vacuum
even {\it before} passing to the VA formulation that eliminates the higher-derivative interactions. This is because of the non-analyticity of the
ModMax Lagrangian, which can be resolved in the Hamiltonian formulation. The VA formulation potentially provides a convenient route to
a similar analysis of the superModMax field equations, and this was another motivation to develop this formulation.

It was mentioned in the Introduction that the ModMax Hamiltonian field equations admit exact plane-wave solutions. Although these
plane waves must be solutions of the superModMax equations if the photino field is set to zero, there is `$0/0$'' ambiguity
because the quadratic fermion term blows up as $S^2+P^2\to 0$. However, preliminary investigations suggest that this difficulty
can also be resolved within a Hamiltomnian formulation, and that the exact plane wave solutions are  half-supersymmetric solutions
of superModMax; we hope to report on this in the near future.

\bigskip

\noindent
{\bf Note added}. While writing this paper we were kindly informed by Sergei Kuzenko that he also has results for a supersymmetric
ModMax electrodynamics and some duality-symmetric superconformal $\mathcal N=2$ models, which will soon appear \cite{Kuzenko:2021cvx}.
\bigskip

\section*{Acknowledgements}
The authors are grateful to Sergey Krivonos, Sergei Kuzenko and especially to Fotis Farakos for very useful discussions.
IB has been partially supported by the Basque Country University program UFI 11/55. Work of IB and DS has also been partially supported  by the Spanish MICINN/FEDER (ERDF EU)  grant PGC2018-095205-B-I00 and by the Basque Government Grant IT-979-16.
PKT has been partially supported by STFC consolidated grant ST/T000694/1.

\appendix
\setcounter{equation}0
\def\theequation{\ref{notation}.\arabic{equation}}
\section{Notation, conventions and relations}\label{notation}
\bea
\eta^{mn}=diag(-,+,+,+)\; , \qquad \Box =-\partial_m\partial^m \; , \qquad \\ \label{sts=}\sigma^m\tilde{\sigma}^n= - \eta^{mn}+\sigma^{mn}= - \eta^{mn} +\frac i 2 \epsilon^{mnpq}\sigma_p\tilde{\sigma}_q\; , \qquad \\
\tilde{\sigma}^m\sigma^n=- \eta^{mn}+\tilde{\sigma}^{mn}= - \eta^{mn}-\frac i 2 \epsilon^{mnpq}\tilde{\sigma}_p\sigma_q\; , \qquad \\
\{{\mathcal D}_{\alpha}, \bar{{\mathcal D}}_{\dot{\alpha}}\}=-2i\sigma^m\partial_m \; ,  \qquad \\ \label{Dal=}
{\mathcal D}_{\alpha}=\partial_\alpha +i(\sigma^m\bar{\theta})_\alpha \partial_m=
e^{-i\theta\sigma^n\bar{\theta}\partial_n}\,\partial_\alpha\, e^{i\theta\sigma^m\bar{\theta}\partial_m}\; , \qquad \\
 \label{bDdal=}\bar{{\mathcal D}}_{\dot\alpha}=({\mathcal D}_{\alpha})^*=-\bar{\partial}_{\dot\alpha} -i({\theta}\sigma^m)_{\dot\alpha}\partial_m =
- e^{i\theta\sigma\bar{\theta}\partial}\, \bar{\partial}_{\dot\alpha}\, e^{-i\theta\sigma\bar{\theta}\partial}\; , \qquad \\
\label{exp=}e^{i\theta\sigma\bar{\theta}\partial}=1+ i\theta\sigma^m\bar{\theta}\partial_m-\frac 1 4 \theta\theta\,\bar{\theta}\bar{\theta}\, \Box\;.
\eea

Other properties of the $\sigma$--matrices are
\begin{subequations}
\begin{align}
\label{s2s+ss2=esA}
\sigma_{ab}\sigma_c+\sigma_c\tilde{\sigma}{}_{ab}=  2\ii\varepsilon_{abcd} \sigma^d \; , \qquad \tilde{\sigma}{}_{ab}\tilde{\sigma}_c+\tilde{\sigma}_c{\sigma}{}_{ab}= -2\ii\varepsilon_{abcd} \tilde{\sigma}^d  \; ,    \quad \\
\label{s2s-ss2=sA}
\sigma_{ab}\sigma_c-\sigma_c\tilde{\sigma}{}_{ab} = -4{\sigma}_{[a}\eta_{b]c} \; , \qquad \tilde{\sigma}{}_{ab}\tilde{\sigma}_c-\tilde{\sigma}_c{\sigma}{}_{ab}=- 4\tilde{\sigma}_{[a}\eta_{b]c}  \; ,    \quad
 \\
\label{sss-sss=esA}
\sigma_{b}\tilde{\sigma}_a\sigma_c- \sigma_c\tilde{\sigma}_a\sigma_b=  -2\ii\varepsilon_{abcd} \sigma^d \; , \qquad \tilde{\sigma}_{b}{\sigma}_a\tilde{\sigma}_c- \tilde{\sigma}_c{\sigma}_a\tilde{\sigma}_b= 2\ii\varepsilon_{abcd} \tilde{\sigma}^d \; .   \quad  \\ \label{s2s2=}
\sigma^{mn}\sigma^{pq}= -2\eta^{m[p}\eta^{q]n} -i\epsilon^{mnpq} + 2\sigma^{m[p}\eta^{q]n}- 2\sigma^{n[p}\eta^{q]m} \qquad
  \end{align}
  \end{subequations}
  The spinor indices are raised and lowered as follows:
\be\label{updown}
\theta_\alpha=\epsilon_{\alpha\beta}\theta^\beta,\qquad \theta^\alpha=\epsilon^{\alpha\beta}\theta_\beta,\qquad (\epsilon^{\alpha\gamma}\epsilon_{\gamma\beta}=\delta^\alpha_\beta)
\ee
and similarly for the dotted indices. The square of the Grassmann variables is defined as
\be
\theta^2=\theta^\alpha\theta_{\alpha}\,,\qquad \bar\theta^2=(\theta^2)^*=\bar\theta_{\dot\alpha}\bar\theta^{\dot\alpha}.
\ee
\\
The following commutators are often useful
\be
{} [{\mathcal D}_\alpha, {\bar{{\mathcal D}}}^2]= -4i \sigma^m_{\alpha\dot\alpha}\partial_m{\bar{{\mathcal D}}}^{\dot{\alpha}}\; , \qquad
 {}[{\mathcal D}^2, {\bar{{\mathcal D}}}^2]= -4i \sigma^m_{\alpha\dot\alpha}\partial_m[ {\mathcal D}^\alpha, {\bar{{\mathcal D}}}^{\dot{\alpha}} ]\; , \qquad
\ee
The component structure of the superfield ${\mathcal D}_\alpha W^\beta$ is
\bea\label{DaWb=}
{\mathcal D}_\alpha W^\beta&=&\delta_\alpha {}^\beta  D + \frac i 2 F_{mn}\sigma^{mn}{}_\alpha {}^\beta
-2\theta_\alpha (\partial_m\bar{\lambda}\tilde{\sigma}^m)^\beta + 2({\sigma}^m\bar{\theta})_\alpha \partial{\lambda}^\beta  \nonumber \\ && -i \theta\sigma_m\bar{\theta}\left(\sigma^{mn}{}_\alpha {}^\beta  \partial_n D + \frac i 2 \partial_n F_{pq}\left(
\sigma^{mn} \sigma^{pq}\right){}_\alpha {}^\beta \right) \nonumber \\
&&  - i \theta\theta\, (\sigma^m\bar{\theta})_\alpha (\partial_m\partial_n \bar{\lambda}\tilde{\sigma}^n)^\beta  - i \theta\theta\, (\sigma^m\bar{\theta})_\alpha (\partial_m\partial_n \bar{\lambda}\tilde{\sigma}^n)^\beta  
+ i \bar{\theta}\bar{\theta}\, \theta_\alpha\Box \lambda^\beta \nonumber \\
&& +\, \frac 1 4  \theta\theta\, \bar{\theta}\bar{\theta} \left(\delta_\alpha {}^\beta \Box D + \frac i 2 \Box F_{mn}\sigma^{mn}{}_\alpha {}^\beta\right). 
\eea
The trace of \eqref{DaWb=} gives the superfield containing the gauge supermultiplet equations of motion and Bianchi identity
\bea\label{DW}
{\mathcal D}_\alpha W^\alpha &=&2 D-2(\theta\sigma^m\partial_m\bar\lambda
+\partial_m\lambda\sigma^m\bar\theta)+\theta\sigma^m\bar\theta\,(2\partial_nF^{n}{}_m-i\epsilon_{nlpq}\partial^lF^{pq})\nonumber\\
&& \ -i(\bar\theta^2\, \theta\Box\lambda-\theta^2 \,\bar\theta \Box \bar \lambda)+\frac 12 \theta^2\bar\theta^2\,\Box D\,.
\eea
When the field strength Bianchi identity is satisfied, ${\mathcal D}W$ is a real linear superfield, i.e. ${\mathcal D}^\alpha W_\alpha=\bar {\mathcal D}_{\dot\alpha}\bar W^{\dot\alpha}$ and ${\mathcal D}^2 {\mathcal D}W=\bar {\mathcal D}^2 {\mathcal D}W=0.$

The superfield ${\mathcal D}^2W^2$ can be alternatively written as
\bea\label{D^2W^2}
{\mathcal D}^2W^2 &=&-2({\mathcal D}^\alpha W^\beta) ({\mathcal D}_\alpha W_\beta)+2W^\alpha {\mathcal D}^2 W_\alpha\nonumber\\
&=& - ({\mathcal D}_\alpha W^\alpha)^2
-2 ({\mathcal D}^{(\alpha} W^{\beta)})({\mathcal D}_{(\alpha} W_{\beta)})+2W^\alpha {\mathcal D}^2 W_\alpha\,.
\eea

\subsection*{\bf Proof of the relation \eqref{L=fM}}

Note that in $f(M)$ we can shift $M$ by any terms proportional to powers of $\Lambda$ and/or $\bar \Lambda$ (where $X$ is some superfield)
\be\label{fM+X}
f(M+\Lambda \cdot X)=f(M)+\frac{\partial f}{\partial M}\, \Lambda\cdot X +O(\Lambda^2).
\ee
This shift of $M$ does not change the l.h.s. of \eqref{L=fM} since the terms proportional to $\Lambda$ and/or $\bar\Lambda$ are annihilated by $\Lambda^2\bar\Lambda^2$, i.e.
\be\label{M+X}
\frac 1{16} \Pi^2\bar\Pi^2\Big(\Lambda^2\bar\Lambda^2\,f(M)\Big)=\frac 1{16} \Pi^2\bar\Pi^2\Big(\Lambda^2\bar\Lambda^2\,f(M+\Lambda \cdot X)\Big)\,.
\ee
Now note that $\frac 1{16} \Pi^2\bar\Pi^2(\Lambda^2\bar\Lambda^2\,M)$ is
a particular form of the shift $M+\Lambda \cdot X$, because of the  identities \eqref{PiLambda}. With this choice equation \eqref{M+X} becomes
\be\label{M+}
\frac 1{16} \Pi^2\bar\Pi^2\Big(\Lambda^2\bar\Lambda^2\,f(M)\Big)=
\frac{1}{16} \Pi^2\bar\Pi^2\left[ \Lambda^2\bar\Lambda^2 f 
\left( \frac{1}{16} \Pi^2\bar\Pi^2\left(\Lambda^2\bar\Lambda^2 M\right)\right) \right]\, . 
\ee
Now notice that in the r.h.s. of \eqref{M+} the (leftmost) derivatives $\Pi$ and $\bar\Pi$ may act only on $\Lambda$ and $\bar\Lambda$; e.g. if $\Pi$ acts on $f(\frac 1{16} \Pi^2\bar\Pi^2(\Lambda^2\bar\Lambda^2\,M))$ the result is zero, because $\Pi^3=0$. Thus, the r.h.s. of \eqref{M+} reduces to the r.h.s. of \eqref{L=fM}.

\if{}
\bibliographystyle{abe}
\bibliography{references}{}
\end{document}
\fi

\providecommand{\href}[2]{#2}\begingroup\raggedright\endgroup

\end{document}